%% file: HHLrep.tex
\documentclass[reqno,12pt]{article}
\usepackage{enumerate}
\usepackage{ifpdf}
\usepackage{ae}
\usepackage[T1]{fontenc}
\usepackage[ansinew]{inputenc}
\usepackage{mathrsfs}
\usepackage{amsmath}
\usepackage{amssymb}
\usepackage{graphicx}
\usepackage{color}
\definecolor{darkblue}{cmyk}{0.9,0.9,0,0}
\definecolor{darkgreen}{rgb}{0,0.55,0}
\usepackage[colorlinks=true,linkcolor=darkblue,citecolor=darkblue,urlcolor=darkblue]{hyperref}
\usepackage{epsfig}

\newcommand{\captionfonts}{\it\small}

\makeatletter
\long\def\@makecaption#1#2{
  \vskip\abovecaptionskip
  \sbox\@tempboxa{{\captionfonts #1: #2}}
  \ifdim \wd\@tempboxa >\hsize
    {\captionfonts #1: #2\par}
  \else
    \hbox to\hsize{\hfil\box\@tempboxa\hfil}
  \fi
  \vskip\belowcaptionskip}
\makeatother

\newcommand{\mincludegraphics}[1]{
\ifpdf
    \includegraphics[width=100mm]{#1}
\else
    \com{PDF-picture replacement}
\fi
}

\newcommand{\beq}{\begin{equation}}
\newcommand{\eeq}{\end{equation}}
\newcommand{\beqq}{\begin{equation*}}
\newcommand{\eeqq}{\end{equation*}}
\newcommand\beqa{\begin{eqnarray}}
\newcommand\eeqa{\end{eqnarray}}
\newcommand\beqaa{\begin{eqnarray*}}
\newcommand\eeqaa{\end{eqnarray*}}
\newcommand\bea{\begin{array}}
\newcommand\eea{\end{array}}

\def\XXint#1#2#3{{\setbox0=\hbox{$#1{#2#3}{\int}$}
\vcenter{\hbox{$#2#3$}}\kern-.5\wd0}}

\newcommand{\nn}{\nonumber}
\newcommand{\com}[1]{(*{\textbf{#1}}*)}

\newcommand{\neqa}{\nonumber\end{eqnarray}}
\newcommand{\la}[1]{\label{#1}}

\newcommand{\eq}[1]{eq.(\ref{#1})}

\newcommand{\Tr}{{\rm Tr}}

\renewcommand{\d}{\partial}

\newcommand{\<}{{\langle}}
\renewcommand{\>}{{\rangle}}

\newcommand{\cL}{{\cal L}}
\newcommand{\bv}{\overline{{\rm v}}}

\newcommand{\re}{\relax{\rm I\kern-.18em R}}

\newcommand{\caB}{{\mathscr B}}

\newcommand{\caA}{{\mathscr A}}

\renewcommand{\sp}{p\hspace{-.40em}/}

\def\su2{{SU(2)}}

\def\[{\left[}
\def\]{\right]}

\def\({\left(}
\def\){\right)}
\def\[{\left[}
\def\]{\right]}

\def\<{\langle}
\def\>{\rangle}

\def\i2{\frac{i}{2}}

\def\bU{{\mathbb U}}

\def\O{{\mathcal O}}

\def\spi{\relax{\rm \pi\kern-0.5em /}}
\def\sA{\relax{\rm A\kern-0.5em /}}
\def\sp{\relax{\rm p\kern-0.5em /}}
\def\sd{\relax{\rm \d\kern-0.5em /}}
\def\sk{\relax{\rm k\kern-0.5em /}}
\def\sn{\relax{\rm n\kern-0.5em /}}
\def\sl{\relax{\rm l\kern-0.5em /}}
\def\sP{\relax{\rm P\kern-0.7em /}}
\def\sBethe{\relax{\rm \Bethe\kern-0.5em /}}

\def\bu{{\bf u}}
\def\bbu{{\bf\bar u}}
\def\bv{{\bf v}}

\def\bU{{\bf U}}

\usepackage{varioref}
\usepackage{makeidx}
\makeindex

\usepackage[english]{babel}
\begin{document}

\thispagestyle{empty}

\renewcommand{\thefootnote}{\fnsymbol{footnote}}
\setcounter{page}{1}
\setcounter{footnote}{0}
\setcounter{figure}{0}
\begin{center}
$$$$
{\Large\textbf{\mathversion{bold}
Tailoring Three-Point Functions and Integrability II \\
Weak/strong coupling match
}\par}

\vspace{1.0cm}

\textrm{Jorge Escobedo$^{a,b}$, Nikolay Gromov$^c$, Amit Sever$^a$ and Pedro Vieira$^{a}$}
\\ \vspace{1.2cm}
\footnotesize{

\textit{$^{a}$ Perimeter Institute for Theoretical Physics\\ Waterloo,
Ontario N2L 2Y5, Canada}  \\
\texttt{jescob,amit.sever,pedrogvieira@gmail.com} \\
\vspace{4mm}
\textit{$^{b}$ Department of Physics and Astronomy \& Guelph-Waterloo Physics Institute,\\
University of Waterloo, Waterloo, Ontario N2L 3G1, Canada} \\
\vspace{4mm}
\textit{$^c$ King's College London, Department of Mathematics WC2R 2LS, UK \& \\
St.Petersburg INP, St.Petersburg, Russia} \\
\texttt{nikgromov@gmail.com}
\vspace{3mm}
}

\par\vspace{1.5cm}

\textbf{Abstract}\vspace{2mm}
\end{center}

\noindent
We compute three-point functions of single trace operators in planar ${\cal N}=4$ SYM. We consider the limit where one of the operators is much smaller than the other two. We find a precise match between weak and strong coupling in the Frolov-Tseytlin classical limit for a very general class of classical solutions. To achieve this match we clarify the issue of back-reaction and identify precisely which three-point functions are captured by a classical computation.

\vspace*{\fill}

\setcounter{page}{1}
\renewcommand{\thefootnote}{\arabic{footnote}}
\setcounter{footnote}{0}

\newpage

 \def\nref#1{{(\ref{#1})}}

\tableofcontents

\section{Introduction}
In this paper we consider the computation of planar three-point functions of single
trace gauge-invariant operators in $\mathcal{N}=4$ SYM. These were studied recently at weak  \cite{Okuyama:2004bd,Roiban:2004va,Alday:2005nd,paper1} and strong coupling \cite{recentpapers1,recentpapers2,Roiban:2010fe}. Here, we will take the classical limit of  these computations and find an exact match.
In the study of the two-point functions related to the spectrum problem, a similar match was pivotal in establishing a first bridge
between weak and strong coupling \cite{FT,ing,Beisert:2003xu,Kruczenski:2003gt,KMMZ}.
 We hope the same to be true for three-point functions. In this introduction we sketch
 the logical flow of the paper and present some of our main results.

First let us define clearly the object we will be interested in. These are the structure
constants $C_{123}$ which appear in the three-point functions
\beq\la{3pt}
\<\O_1 (x_1) \O_2 (x_2) \O_3 (x_3) \> = \frac{1}{N} \frac{C_{123}}{|x_{12}|^{\Delta_1+\Delta_2-\Delta_3} |x_{23}|^{\Delta_2+\Delta_3-\Delta_1} |x_{31}|^{\Delta_3+\Delta_1-\Delta_2}} \, ,
\eeq
where we normalize the two-point functions as $\< \O_i(x) \bar \O_i(0)\>=|x|^{-2\Delta_i}$.
They are not completely unambiguous because we can still multiply $\O_i$ by a phase. That would not affect the
normalization of the two-point function, but it would change $C_{123}$ by that phase.\footnote{The phase is important for bootstrapping higher-point functions. This goes beyond the scope of the current analysis.}
On the other hand the absolute value $|C_{123}|$ is well-defined and is the object we will be computing. To remove the dependence on the normalization convention, we will also consider the ratio
\beq\la{rfunction}
r = \left| \frac{C_{123}}{C_{123}^{\circ\circ\circ}}\right| \, .
\eeq
In this expression $C_{123}^{\circ\circ\circ}$ is the correlation function between three
chiral primary operators $\O_i^{\text{BPS}}$ with the same charges as $\O_i$. The three-point
function $C_{123}^{\circ\circ\circ}$ is known \cite{Lee:1998bxa} and are
non-renormalized. Hence $r$ contains as much information as $C_{123}$ itself
but contains no trivial cumbersome combinatoric factors.

At weak coupling, the leading order planar computation is a purely combinatorial problem.
At 't Hooft coupling $\lambda=0$ there is a large number of operators with the same classical
dimension. Therefore we need to perform degenerate perturbation theory. That is, we must use
the eigenvectors of the one-loop dilatation operator that lift the degeneracy \cite{Beisert:2002bb}. For example,
for operators made out of two complex scalars $Z$ and $X$ we have
\beq
\mathcal{O}(x) = \sum_{1\le n_1<\dots <n_N \le L} \psi(n_1, \dots,n_N) \Tr\Big( \overbrace{\!\underbrace{Z\dots Z}_{n_1-1 \text{ fields}} X Z \dots Z}^{n_2-1\text{ fields}} X Z \dots\Big) \la{operators}
\eeq
where
\beqa
\psi(n_1)&=&\phi_1\,, \nn \\
\psi(n_1,n_2)&=&\phi_{12} \,\,+ \mathcal{S}_{21} \,\phi_{21}\,, \nn \\
\psi(n_1,n_3,n_3)&=& \phi_{123}+ \mathcal{S}_{21}\, \phi_{213}+ \mathcal{S}_{32} \, \phi_{132}+\mathcal{S}_{31}\mathcal{S}_{23} \,\phi_{312}+\mathcal{S}_{31}\mathcal{S}_{21} \,\phi_{231}+\mathcal{S}_{32}\mathcal{S}_{31} \mathcal{S}_{21} \,\phi_{321}\,,\nn
\eeqa
and so on.
In these expressions $\phi_{rs\dots}=e^{i p_r n_1+i p_s n_2+\dots}$ and
$\mathcal{S}_{rs}=\frac{\cot({p_r}/{2})-\cot({p_s}/{2})+2i}{\cot({p_r}/{2})-\cot({p_s}/{2})-2i}$.
For general $N$ we have $N!$ terms. We can think of the scalars $X$ as particles with momenta $p_r$
propagating in the $Z$ vacuum. They interact with each other in a purely pairwise way. Each scattering
event is governed by the two particle S-matrix $\mathcal{S}_{rs}$. These integrable wave functions $\psi(n_1,\dots,n_N)$
go by the name of Bethe wave functions. The momenta $p_r$ are those of particles in a finite ring (of length equal to
the total number of elementary operators in the single trace). Hence they are quantized. The corresponding
quantization equations that follow from the periodicity of the wave functions are the so-called Bethe
equations.
One often thinks of the states (\ref{operators})
as states in a quantum spin chain \cite{Minahan:2002ve,Staudacher:2004tk}. In this picture we think of the $Z$ fields as spins up
forming a ferromagnetic vacuum while the $X$ fields are spins down called magnons.
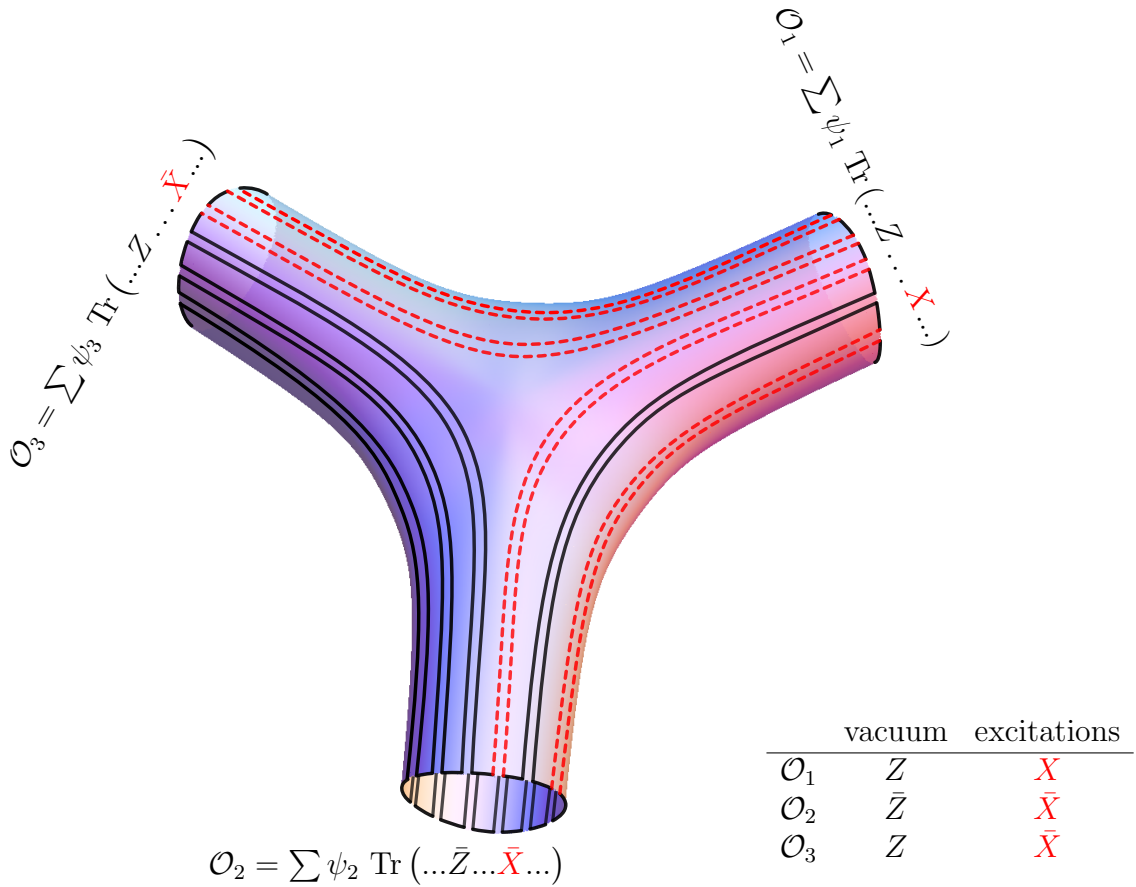
\begin{figure}[t]
\centering
\def\svgwidth{12cm}

\ifpdf
    \input{3D3P.pdf_tex}
\else
    \com{PDF-picture replacement}
\fi

\caption{(a) Three-point function of $SU(2)$ operators at tree level considered in \cite{paper1}.
All contractions are such that R-charge is preserved.  This is the simplest non-trivial configuration
which is not extremal.
}\label{3ptfunction}
\end{figure}

The tree-level problem is then trivial to enunciate (although very hard to solve in full generality):
we need to pick three operators like (\ref{operators}) with different kinds of scalars, as represented in figure \ref{3ptfunction}.
Then we need to sum over all possible Wick contractions of these three operators as represented in the same figure. This problem is complex because the number of terms in the Bethe wave functions grows factorially. The final form of the result typically involves determinants of the Hessian of the Yang-Yang functional\footnote{The Yang-Yang functional is a functional whose minimization leads to the Bethe equations.} and sums over all possible ways of partitioning the momenta $p_j$ into two sets. They are quite involved; see
table 2 of \cite{paper1}.\footnote{Needless to say, the example we just described is not the most general one.
A general single trace operator can contain any of the six scalars, four covariant derivatives and sixteen fermions
of $\mathcal{N}=4$ SYM. The operators are then described by a Nested Bethe ansatz and the tools needed to efficiently
solve the weak coupling combinatorial problem are not yet developed. For example, the main ingredients in the
combinatorial computation are scalar products between Bethe states. These scalar products are known
only for $SU(2)$ case. We are currently exploring this problem \cite{toappear2}.}

To make analytic progress at weak coupling it is useful to consider some simplifying limits. In this paper we will consider a classical limit where two of the three operators are very large while the third operator is small (compared to the other two). This limit will allow us to obtain simple analytic results for the structure constants at weak coupling. Also, and this is the main result of the current paper, it will enable the direct comparison with the strong coupling results.

\begin{figure}[t]
\centering
\def\svgwidth{18cm}

\ifpdf
    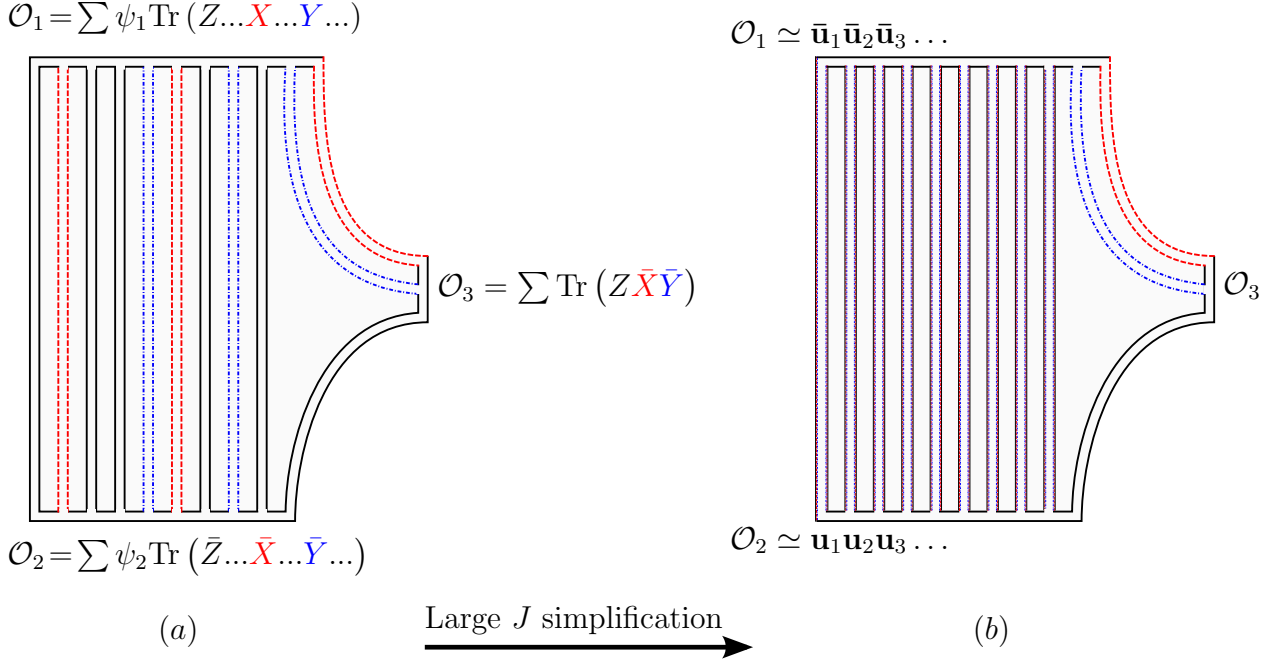
\else
    \com{PDF-picture replacement}
\fi

\caption{The setup considered in this paper for three-point function of $SU(3)$ operators. Two operators are large and roughly conjugate to each other while the third operator is small. (a) Exact setup. The tree-level contraction of three pure eigenstates \cite{paper1}. These are very entangled states with order $(\text{number of $X$'s and $Y$'s})!$ terms. All contractions are such that R-charge is preserved. (b) Classical limit. The tree-level contraction of the small operator with the corresponding two large coherent states. The coherent states are not entangled at all, so their contraction is trivial.
}\label{HHL}
\end{figure}

In this paper we consider a generalization of the setup of figure \ref{3ptfunction} studied in \cite{paper1}, namely we consider $SU(3)$ operators made out of $SU(3)$ complex scalars as indicated in (\ref{Vacchoice2}), see  figure \ref{HHL}a.
\beq\la{Vacchoice2}
\begin{array}{cccll}
& \text{vacuum} & \text{excitations} && \text{notations}\\ \hline
\O_1 & Z & X \text{ and } Y && \#\{X,Y,Z\}\,= \{J_1,J_2 ,J_3\} \\
\O_2 & \bar Z & \bar  X \text{ and } \bar Y && \#\{\bar X,\bar Y,\bar Z\}= \{J_1-j_1,J_2-j_2 ,J_3+j_3\}  \\
\O_3 & Z & \bar X \text{ and } \bar Y && \#\{\bar X,\bar Y,Z\}= \{j_1,j_2 ,j_3\}
\end{array}
\eeq
The total length of operators $\O_1$ and $\O_3$ are, respectively,
$$J\equiv J_1+J_2+J_3 \qquad \text{and} \qquad j\equiv j_1+j_2+j_3\,.$$
The operator $\mathcal{O}_3$ is taken to be small, that is $j  \ll J$. For comparison with strong coupling, we will focus on the case where $\mathcal{O}_3$ is a protected chiral primary operator. At weak coupling, this condition can be trivially relaxed, see section \ref{nonBPS}.

The operators $\O_1$ and $\O_2$ are taken to be very large and with a lot of excitations such that their density is kept fixed in the large length limit. Moreover, we only consider low lying excitations around the ferromagnetic vacuum. These are excitations with long wavelength. That is, the momenta of the individual excitations of operators  $\O_1$ and $\O_2$ are of order $1/J$.
The energy of the individual excitations is of order $1/J^2$ so that the total energy is of order $1/J$. In short, we will consider the Sutherland classical limit \cite{Sutherland:1995zz}, rediscovered in the AdS/CFT context in \cite{ing}.

The operator $\mathcal{O}_1$ and $\mathcal{O}_2$ are taken to be \textit{roughly} the complex conjugate of each other in a sense that will be made precise later. The physical picture is that the large operator $\O_2$ interacts with the small operator $\mathcal{O}_3$ to become the slightly modified final large operator $\O_1$. Note that we can not take $\O_1$ to be \textit{exactly} the conjugate  of $\O_2$. This would lead to a vanishing three-point function by R-charge conservation.

Now, in the classical limit the excitations have long wavelength and can be depicted as changing the spin configuration slowly as one goes around the spin chain. Magnon dynamics, in this limit, are well approximated by the Landau-Lifshitz field theory \cite{LL,Kruczenski:2003gt}. More explicitly, we can use coherent states to approximate the exact states (\ref{operators}) or rather their Nested Bethe Ansatz generalization. The coherent states mimic the exact Bethe states in the following precise sense: when computing with them an average of a classical quantity such as the total spin or energy, they yield the same result up to finite size corrections. In the spin chain language,
\beq\la{classical}
\frac{\<\psi_1|\O_\text{classical}|\psi_1\>}{\<\psi_1|\psi_1\>}=\frac{\<\varphi_1|\O_\text{classical}|\varphi_1\>}{\<\varphi_1 | \varphi_1\>}\(1+O(\tfrac{1}{J})\)
\eeq
where $\psi_1$ denotes the exact Bethe state while $\varphi_1$ denotes the coherent state. More explicitly, the coherent states are simply
\beqa
\O_1(x)=\dots ({\bf u}^{(n)} \cdot \Phi) ({\bf u}^{(n+1)} \cdot \Phi) \dots \dots\la{cohop}
\eeqa
where $\Phi=(X,Y,Z)$ and ${\bf u}^{(n)}=(u_1^{(n)},u_2^{(n)},u_3^{(n)})$ is a complex vector which is unitary and slowly varying, i.e.\ ${\bf u}^{(n)}\cdot \bar {\bf u}^{(n)}=1$ and ${\bf u}^{(n+1)}-{\bf u}^{(n)} = \mathcal{O}(1/J)$. We can approximate the vector ${\bf u}^{(n)}$ by a continuous field ${\bf u}(\sigma)$ where $\sigma = 2\pi n/J$ is a continuous parameter taking values between $0$ and $2\pi$. This continuous field obeys the Landau-Lifshitz equations of motion described below.

Now, the key observation is that a small operator $\O_3$ is roughly a classical operator inserted between operators $\O_1$ and $\O_2$, see also \cite{Roiban:2004va}. Note that this is only true provided $j \ll J$ which is the limit we consider. Then in the computation of the Wick contractions between the three operators, one can simply replace the large operators by the corresponding coherent states. This simplifies the computation enormously. Instead of $N_1!$ terms in the Bethe state we have a single term in the coherent state description! The coherent state is not entangled at all and the Wick contractions are trivial, see figure \ref{HHL}b. The computation is described in detail in section \ref{coherentderivation}
and the result is almost obvious, we find\footnote{We are doing quantum mechanics so each state in the Hilbert space is a ray. This translates into a $U(1)$ gauge symmetry of individually
multiplying the $\bu^{(j)}$'s by a phase. Strictly speaking (\ref{main}) is written in a conformal-like gauge introduced in Sec.\ref{coherentderivation}.}
\beq
r=\left|\frac{C^{\bullet\bullet\circ}_{123}}{C_{123}^{\circ\circ\circ}} \right|= \left| \frac{1}{ v_1^{j_1} v_2^{j_2}\,\bar v_3^{j_3}}
\int_0^{2\pi} \!\! \frac{d\sigma}{2\pi}\, u_1^{j_1} u_2^{j_2}\,\bar u_3^{j_3} \right|\, \la{main}
\eeq
where  $v_i =\sqrt{{ {J_i}}/{J}}$.
The notation ${C^{\bullet\bullet\circ}_{123}}$ emphasizes that we compute the correlation function where operators $\O_1$ and $\O_2$ are non-protected while $\O_3$ is protected.
As anticipated above, the classical limit allowed us to dramatically simplify the exact quantum results.

We will now switch gears and move to strong coupling.
We will  show that in a particular limit the strong coupling computation precisely reproduces our main result (\ref{main}). At strong coupling each gauge-invariant operator is described by a single string: the large operators are described by classical strings, while the small operator is a
quantum string. Since we consider operators made out of scalars only, the string motion is non-trivial in $S^5$ and is point-like in $AdS_5$. The simplest classical string solutions in $S^5$ are point-like strings rotating around one of the equators,\footnote{We will often use the same letter for quantities arising at strong coupling and quantities appearing at weak coupling. We do so for quantities that will later be matched between weak and strong coupling. We hope this will not raise any confusions.}
\beq
{\bf U } = e^{i \kappa \tau} {\bf v} \la{BMNstring} \, ,
\eeq
where ${\bf U }=\{X_1+i X_2, X_3+i X_4, X_5+i X_6\}$ and $X_i$ are the sphere embedding coordinates. The latter square to $1$ so that ${\bf U}$ is unitary, ${\bf U } \cdot \bar{\bf U }=1$. In (\ref{BMNstring}), ${\bf v}$ is a constant unitary vector which can be parametrized as
\beq\la{BPS}
v_a = \sqrt{\frac{J_a}{J}} \quad , \quad J=J_1+J_2+J_3 \,,
\eeq
where $J=\sqrt{\lambda}\,\kappa$ is the total angular momentum of the point-like string and $J_i$ is the angular momentum in the $(X_{2i-1},X_{2i})$ plane. The energy of this solution is also equal to $E=\sqrt{\lambda}\,\kappa$. In the classical limit we should have $\kappa \gg 1$. This solution is simply the BMN point-like string \cite{BMN} dual to the BPS operator
\beq
\mathcal{O}_{\text{BPS}} \propto   {\rm Tr} \( X^{J_1} Y^{J_2} Z^{J_3}\) \,+\, \text{all possible permutations} \, .
\eeq
A simple generalization of the BMN solution is given by considering solutions of the form
\beq
{\bf U } = e^{i \kappa \tau} {\bf u} \, ,\la{generalstring}
\eeq
where $\kappa\to\infty$ while $\kappa\,\d_{\tau} {\bf u}$ is held fixed \cite{Kruczenski:2003gt,Stefanski:2004cw}. This is the same \cite{Kruczenski:2003gt} as the famous Frolov-Tseytlin limit \cite{FT} which corresponds to $\lambda,J\to \infty$ with $\lambda/J^2 \ll 1$. For example, the energy of these solutions $E=\sqrt{\lambda}\,\kappa$ is given by
\beq
E = J\[1 + \frac{\lambda}{2J^2}\int_0^{2\pi} \frac{d\sigma}{2\pi}\;\d_\sigma\bar {\bf u}\cdot\d_\sigma {\bf u}+\dots\]\;. \la{resultstrong}
\eeq
One of the remarkable features of this expansion is that an expansion in powers of $\lambda/J^2$ resembles a weak coupling expansion in $\lambda$. Indeed (\ref{resultstrong}) turns out to coincide precisely with the weak coupling spin chain spectrum in the classical limit described by the coherent states introduced above! This limit was instrumental in establishing a first bridge between weak and strong coupling for non-protected operators in the spectrum problem \cite{FT,ing,Beisert:2003xu,Kruczenski:2003gt,KMMZ}.

We will now show that for three-point functions this limit is also extremely useful. It allows us to match the strong coupling results with the weak coupling prediction (\ref{main}) in a straightforward way.\footnote{The possibility of obtaining a match for the structure constants by performing this limit was first pointed out in \cite{paper1,recentpapers3}} Let us give a flavor of how it works, the details are presented in section \ref{sectionstrong}. At strong coupling, Zarembo computed the three-point function of two large classical operators and a small BPS operator \cite{recentpapers1}. His result leads to
\beq
r=\left|\frac{C^{\bullet\bullet\circ}_{123}}{C_{123}^{\circ\circ\circ}} \right|= \left| \mathcal{R} \, \frac{\sqrt{\lambda} /J}{v_1^{j_1} v_2^{j_2} \bar v_3^{j_3} } \int\limits_{-\infty}^\infty d\tau_e
\int\limits_0^{2\pi} \frac{d\sigma}{2\pi}
\frac{U_1^{j_1}U_2^{j_2}\bar U_3^{j_3}}{\cosh^j\kappa\tau_e}\[
\frac{2\kappa^2}{\cosh^2\kappa\tau_e}
-\kappa^2-\d_a {\bar{\bf U}}\cdot\d_a {\bf U}
\] \right|
\la{zarembo}
\eeq
\begin{figure}[t]
\centering
\def\svgwidth{14cm}

\ifpdf
    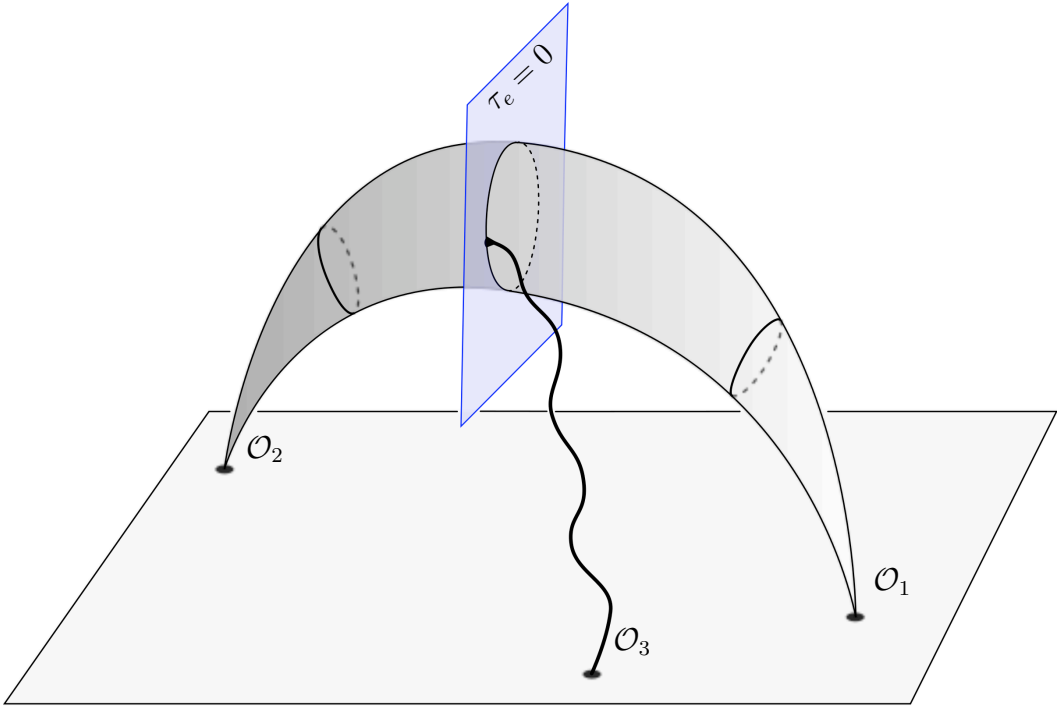
\else
    \com{PDF-picture replacement}
\fi

\caption{The Landau-Lifshitz field theory arises naturally both at weak and
strong coupling. However, this does \textit{not} mean that a match of
the weak and strong coupling results for the structure constants is
guaranteed.  Indeed, the prescriptions for computing
these quantities are a priori very different. The weak coupling overlap of the coherent states results in a single integral over the position of insertion
of the small operator, see (\ref{main}). The
strong coupling computation involves a two-dimensional integration of
a boundary-to-bulk propagator over the full string worldsheet. In the
Frolov-Tseytlin limit the integration becomes localized in the slice
$\tau_e=0$ and only the integral over $\sigma$ survives. In this way
the two computations are matched. Note that $\tau_e=0$ is the only point that is shared between the Lorentzian and Euclidean solutions; the localization mechanism tell us to take a snapshot of the string at precisely this physical point.
}\label{HHL2}
\end{figure}
where $j=j_1+j_2+j_3$ and
\beq
\mathcal{R}=\frac{j +1}{2^{j +2}}\, { j \choose j_3 }  \, .
\la{R}
\eeq
Here, the $\bU$'s represent the big classical state (\ref{generalstring}) and $\bv$ is given by  (\ref{BPS}). For a more detailed description of the several symbols in (\ref{zarembo}) see section \ref{sectionstrong}. For large $\kappa$ only the first term inside the square brackets in (\ref{zarembo}) contributes, the other two are sub-leading as can be shown using the Virasoro constraints. Also, when using (\ref{generalstring}) we can set $\tau=0$ in $\bf u$ since the $1/\cosh^{j+2}(\kappa\tau_e)$ factor localizes the integral at $\tau_e \simeq 1/\kappa$ whereas ${\bf u}$ changes slowly, see figure \ref{HHL2}. Hence the integral over $\tau_e$ factorizes and can be done trivially. In this way we obtain precisely (\ref{main})!
This match is valid for \textit{any} solution in the Frolov-Tseytlin limit. 

In the spectrum problem the generic match of the spectrum problem in the Frolov-Tseytlin limit was extremely useful, see \cite{Beisert:2010jr} for an extensive review. For example, the weak/strong coupling match for the spectrum can be derived by showing that both weak and strong coupling classical limits are properly described by some algebraic curves \cite{KMMZ,BKSZ} which become identical in the Frolov-Tseytlin limit. Such algebraic curves were one of the main ingredients involved in conjecturing the form of the strong coupling Arutyunov-Frolov-Staudacher equations \cite{AFS} and even the all-loop Beisert-Staudacher
equations \cite{BS} which are important input for the Y-system~\cite{Gromov:2009tv}. Those quantum equations can be interpreted as a proper discretization of the classical algebraic curves.\footnote{This discretization technique lead to a vast number of conjectures for all loop Bethe equations for other AdS/CFT systems \cite{ABJM,kostya}.} We hope that the match we are observing for structure constants
could be equally important in the study of three-point functions in $\mathcal{N}=4$ SYM. The next step would be to address the question of the discretization of the classical results and then conjecture the all-loop result for the structure constants, at least for operators that are asymptotically large.

In the rest of this paper we will give a more detailed explanation of the arguments above. We will then discuss some generalizations and finish with some comments and speculations. This paper is organized as indicated in the table of contents.

\section{Strong coupling} \la{sectionstrong}

In this section we give a short review of classical strings in $AdS_5\times S^5$ in the Frolov-Tseytlin limit. We then present a slight generalization of the three-point function of two large classical operators and one small BPS operator considered in \cite{recentpapers1}.

In $AdS_5$ we consider trivial point-like solutions sitting in the middle of global $AdS$ with global time
\beq
t= \kappa \tau \, .
\eeq
On the sphere we consider a generic classical motion. The sphere can be described by three complex embedding coordinates
\beq
(U_1,U_2,U_3)=\bU(\sigma,\tau)
\eeq
constrained by $\bar\bU\cdot\bU=1$.
The equations of motion and the Virasoro constraints are given by\footnote{We use the signature $-+$ i.e. $\d^a\d_a= -\d_\tau^2+\d_\sigma^2$.}
 $\[\d^a\d_a+ (\d^a\bar\bU\cdot\d_a\bU)\,\]U_i=0
$ and $(\d_\tau \bar\bU\pm\d_\sigma \bar\bU)\cdot
(\d_\tau\bU\pm\d_\sigma\bU)=\kappa^2
$ (see for example \cite{Beisert:2004ag}).
An important subclass of classical solutions, which can be
mapped to coherent spin chain states at weak coupling, can
be defined by introducing new coordinates
\beq
\bU(\sigma,\tau)=e^{i\kappa\tau}\bu(\sigma,\tau)\; \la{LLform}
\eeq
and considering the limit \cite{Kruczenski:2003gt}
\beq
\kappa\to\infty  \,\text{  with  } \,  \kappa\,\d_\tau {\bf u}   \,,\,\d_\sigma{\bf u}   \,\text{  held fixed} \,.
\eeq
In this limit the equation of motion and the Virasoro constraints reduce
to
\beq\la{LLeom}
2i\kappa\,\d_\tau {\bf u}=\d_\sigma^2 {\bf u}+2{\bf u}\, (\d_\sigma \bar {\bf u}\cdot\d_\sigma {\bf u})
\eeq
and
\beq
\bar {\bf u}\cdot\d_\sigma {\bf u} = 0\;.
\la{virasoro}
\eeq
The global charges read
\beqa
J_a=\sqrt{\lambda}\int_0^{2\pi} \frac{d\sigma}{2\pi}\;
\(\kappa\bar u_a u_a-i
\bar u_a \d_\tau u_a\)\;.
\eeqa
Using $J=J_1+J_2+J_3$, this equation leads to (\ref{resultstrong})
where the energy of the solution is $E=\kappa\sqrt\lambda$. As explained in detail in the introduction this limit is of great significance for establishing a bridge between weak and strong coupling results.

On the other hand, from a more modern perspective, we now know that the match in the Frolov-Tseytlin limit is, to a great extent, a fortunate accident.\footnote{By taking a certain decoupling limit, it was argued in \cite{Harmark:2008gm} that the spectrum on both sides of the $AdS_5 /CFT_4$ correspondence should match at least at the one-loop level. It would be interesting to see if the same argument holds for structure constants.} A priori the limit $\lambda,J\to \infty$ with $\lambda/J^2 \to 0$ is not the same as $\lambda \to 0$.
Indeed, the match of the Frolov-Tseytlin limit in the spectrum problem breaks down at some loop order in $\mathcal{N}=4$ SYM.\footnote{A three loop mismatch was reported in 
\cite{0510080}.
In more recent
 examples of AdS/CFT dualities it fails already at leading order \cite{ABJM}.
 However the match can be usually achieved after a universal redefinition of the coupling constant. } Still, hopefully the first few orders in the computation of structure constants will match as well. Of course, given the obvious order of limits issue, a more honest way of interpreting this match is as a guiding principle in the pursuit of the all loop result.

\subsection{Holographic three-point functions}
Following Zarembo \cite{recentpapers1}, the holographic three-point
function of two heavy operators and a light chiral primary operator, see figure \ref{HHL}, is given by
\beq\la{gen}
C^{\bullet\bullet\circ}_{123}= \mathcal{C}  \int\limits_{-\infty}^\infty d\tau_e
\int\limits_0^{2\pi} \frac{d\sigma}{2\pi}
\frac{\sqrt\lambda}{J}\frac{U_1^{j_1}U_2^{j_2} \bar U_3^{j_3}}{\cosh^j\kappa\tau_e}\[
\frac{2\kappa^2}{\cosh^2\kappa\tau_e}
-\kappa^2-\d_a {\bar{\bf U}}\cdot\d^a {\bf U}
\] \, ,
\eeq
where
\beq
\mathcal{C} = \frac{\mathcal{R} \, C^{\circ \circ \circ}_{123}}{v_1^{j_1} v_2^{j_2} \bar v_3^{j_3}}\quad \;\;,\;\; \quad
\mathcal{R}=\frac{j +1}{2^{j +2}}\, { j \choose j_3 }  \,
\la{C}
\eeq
and $\textbf{v}$ is associated with the point-like BPS string, see \eqref{BPS}, while $\textbf{U}$ are the three complex sphere embedding coordinates given in \eqref{generalstring}. Finally in our limit
\beq
C_{123}^{\circ\circ\circ}=
Jv_1^{j_1}v_2^{j_2}\bar v_3^{j_3}(j_1+j_2)!\sqrt{\frac{j\, j_3!}{j_2! j_1!j!}} \la{bpssimp}
\eeq
is the structure constant for three chiral primaries with the same charges as the classical operators \cite{Lee:1998bxa}, see appendix \ref{BPSap}.
Note that we are working in Euclidean time $\tau_e=i\tau$ and $j_1,j_2,j_3$ are respectively the number of $\bar X,\bar Y,Z$ fields in operator $\mathcal{O}_3$, such that $j={j_1+j_2+j_3}$. The result (\ref{gen}) is a trivial generalization of \cite{recentpapers1} to a generic $SU(3)$ BPS operator. Moreover, our normalization convention is slightly different.

Given that we are interested in the Frolov-Tseytlin limit, where $\kappa\to\infty$ while $\kappa\,\d_{\tau} {\bf u}$ is held fixed, the integral over $\tau_e$ in \eqref{gen} can be easily computed because only the first term in square brackets contributes at leading order, while the remaining two are subleading, as can be shown using the Virasoro contraint \eqref{virasoro}:
\beqq
\frac{2\kappa^2}{\cosh^2\kappa\tau_e}
-\kappa^2-\d_a {\bar{\bf U}}\cdot\d^a {\bf U}=
\frac{2\kappa^2}{\cosh^2\kappa\tau_e}
-2\d_\sigma\bar{\bf u}\cdot\d_\sigma{\bf u} \,  \overset{\kappa \to \infty}{\simeq} \, \frac{2\kappa^2}{\cosh^2\kappa\tau_e} \, .
\eeqq
Then, using $U_i(\sigma, \tau_e)= e^{\kappa \tau_e} u_i(\sigma, \tau_e)$ from \eqref{generalstring} in \eqref{gen}, we notice that we can set $\tau_e=0$ in $u_i$ because the factor $1/\cosh^{j+2} (\kappa \tau_e)$ localizes the integral around $\tau_e \simeq 1/\kappa$, while $\textbf{u}$ is a slowly changing variable. Hence, we can evaluate the integral over $\tau_e$ as (here $T=\kappa \tau_e$)
\beqq
\int\limits_{-\infty}^{\infty} dT \, \frac{e^{(j_1+j_2-j_3) T}}{\cosh^{j+2}T} = \frac{2^{j+1}}{(j+1) {j \choose j_3}} = \frac{1}{2 \mathcal{R}}\, .
\eeqq
Therefore, the three-point function \eqref{gen} takes the simple form
\beq
C^{\bullet\bullet\circ}_{123}=\left.
J \frac{(j_1+j_2)!}{j_1! j_2!}\sqrt{j\frac{j_1! j_2!j_3!}{j!}}
 \int\limits_0^{2\pi} \frac{d\sigma}{2\pi} \,  u_1^{j_1} \, u_2^{j_2} \, \bar u_3^{j_3} \right|_{\tau_e=0}\, .
 \la{mainstrong}
\eeq
Moreover, if we use the equation that relates $\mathcal{R}$ and $\mathcal{C}$ \eqref{C}, we find a simple expression for the ratio $C^{\bullet\bullet\circ}_{123}/C^{\circ\circ\circ}_{123}$:
\beq
r=\frac{C^{\bullet\bullet\circ}_{123}}{C^{\circ\circ\circ}_{123}} = \left.  \frac{1}{{v_1^{j_1} \, v_2^{j_2} \, \bar v_3^{j_3}}} \int\limits_0^{2\pi} \frac{d\sigma}{2\pi} \,  u_1^{j_1} \, u_2^{j_2} \, \bar u_3^{j_3} \right|_{\tau_e=0}\, .
\la{ratiostrong}
\eeq

\section{Weak coupling}\la{Weak_coupling}

In this section we will describe the weak coupling computation of the structure constants in the limit where two of the operators are large classical operators whereas the third is a small operator. We start by a very representative example and move to the general treatment later on.

\subsection{$SU(2)$ Folded string warm-up} \la{ex1}
Before moving to the general setup, we provide in this section a most convincing evidence for the validity of the weak/strong coupling match. We consider the correlation function between the following three operators. For the small operator $\O_3$ we take the BPS operator
\beq
\O_3 = 2\,\Tr(ZZ\bar X\bar X) + \Tr(Z\bar XZ\bar X) \, .
\eeq
That is we consider $j_1=j_3=2$ and $j_2=0$.  For the classical states $\O_1$ and $\O_2$ we will take operators dual to the folded string with unit mode number. The folded string in the Frolov-Tseytlin limit is given by \cite{FT,Kruczenski:2003gt}\footnote{Coherent states and folded strings have been studied in the past in connection with string splitting and joining in \cite{Casteill:2007td}.}
\begin{align}
u_1 (\sigma,\tau_e) =  e^{ \frac{2(1-q)\text{K}(q)^2}{\pi^2 } \frac{\tau_e}{\kappa}} \,  \sqrt{q} \, \text{sn} \( \frac{2 \text{K}(q)}{\pi}  \sigma |q\) \,\,\, , \,\,\, \bar u_3 (\sigma,\tau_e) =  e^{\frac{2q \text{K}(q)^2}{\pi^2 } \frac{\tau_e}{\kappa}} \,   \text{dn}\( \frac{2   \text{K}(q)}{\pi} \sigma |q\) \, ,
\la{usfolded}
\end{align}
for filling fraction\footnote{For completeness, the energy of this solution is given by $\sqrt{\lambda}\kappa=E=J+{2 \text{K}(q) (\text{E}(q)-(1-q) \text{K}(q))}/{\pi^2 J}$. We will not make use of this expression.}
\beq
\alpha\equiv \frac{J_1}{J_1+J_3}=1- \frac{\text{E}(q)}{\text{K}(q)} \la{alpha} \, .
\eeq
We are in the $SU(2)$ setup of \cite{paper1} which amounts to setting $J_2=0$.
We use Mathematica's conventions where $\text{E}(q)= \,\, $\verb"EllipticE[q]",  $\text{K}(q)=\,\,$\verb"EllipticK[q]", $\text{sn}(x|y)=\,\,$\verb"JacobiSN[x,y]" and $\text{dn}(x|y)=\,\,$\verb"JacobiDN[x,y]".

\begin{figure}[t]
\centering
\def\svgwidth{14cm}
\mincludegraphics{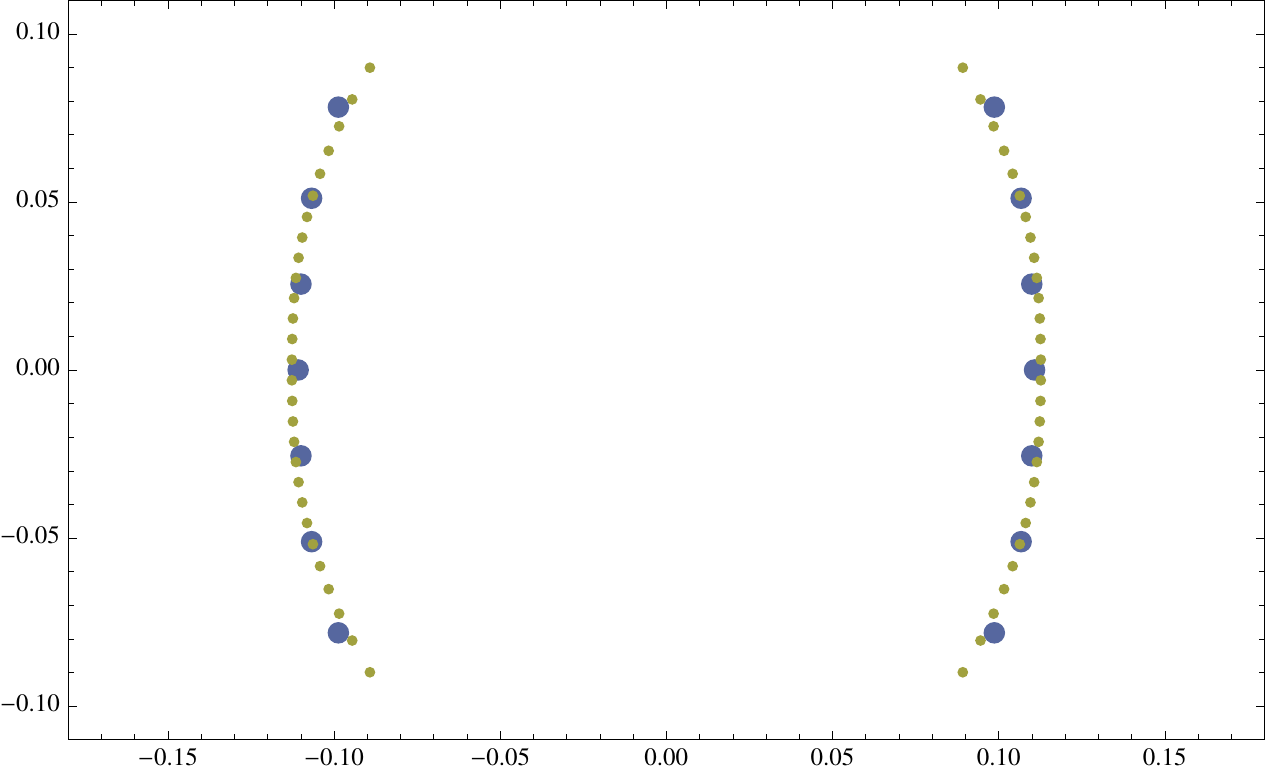}
\caption{Two Bethe roots configurations for a folded string with $\alpha=1/3$ and $J=42,168$, represented by the larger and smaller bullets respectively. The horizontal and vertical axes are the real and imaginary part of $u/J$, where $u$ is the rapidity of an excitation, related to its momentum by $u=\frac{1}{2} \cot \frac{p}{2}$. The case $J=42$ corresponds to the maximum number of roots that we used to compute $r$ with the formulas of \cite{paper1}. Even though this number of roots was not very large, we see that they lie nicely along the cuts formed by the clearly classical configuration $J=168$.
}
\label{foldedcuts}
\end{figure}

\begin{figure}[t]
\centering
\def\svgwidth{14cm}
\includegraphics{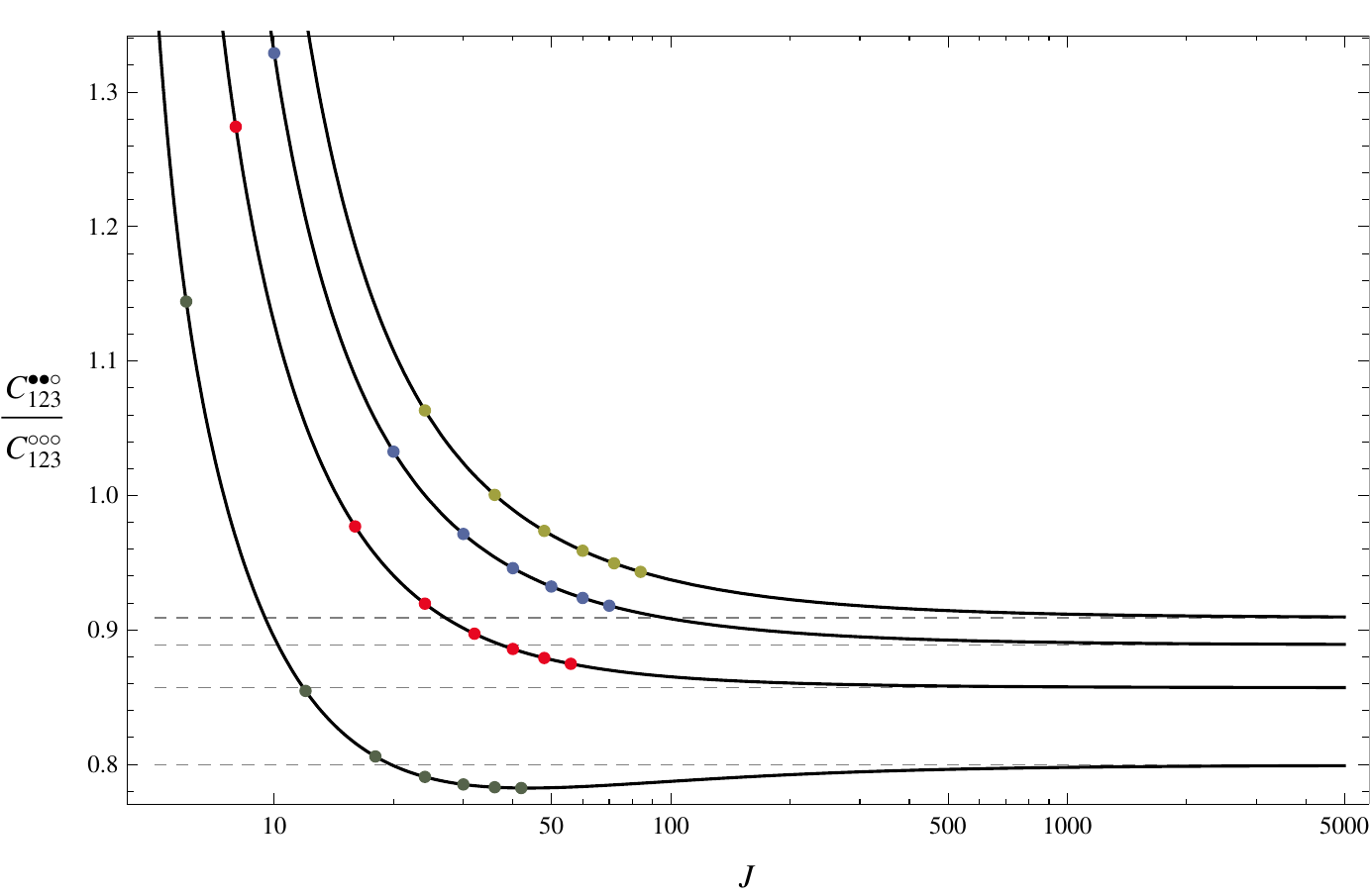}
\caption{Fits of the numerical data (black curves) compared to the analytical prediction (dashed gray lines)
for different values of the filling fraction of operator $\O_1$: $\alpha=1/3,1/4,1/5,1/6$, from bottom to top.
}\label{numplot}
\end{figure}

Plugging these expressions into (\ref{ratiostrong}) we obtain\footnote{For a general BPS operator $\O_3$ in the $SU(2)$ sector, that is with $j_2=0$, we find
\begin{align}
r=  \frac{ \sqrt{\pi} \, \Gamma\(\frac{1+j_1}{2}\) \, q^{j_1/2}\, (1-q)^{j_3/2} \, _2F_1 \(\frac{1}{2},\frac{1+j}{2}; \frac{2+j_1}{2};q\) }{\alpha_1^{j_1/2} \, \alpha_3^{j_3/2} \, j_1 \, \Gamma\(\frac{j_1}{2}\) \, \text{K}(q) } \, ,
\la{rfolded}
\end{align}
where $j_1$ is an even number so that when $\O_2$ is a zero-momentum state, so is $\O_1$.}
\begin{align}
r=  \frac{\pi \, q\, (1-q) \, _2F_1 (\frac{1}{2},\frac{5}{2};2;q)}{4 \,  \alpha (1-\alpha)  \, K(q) } \, .
\la{intsigma}
\end{align}
For different filling fractions $\alpha$ we find $q$ from (\ref{alpha}), plug it in this expression and get a number. For example, for $\alpha=1/4$ we find
\beq
r\simeq 0.856897\, .
\la{1/3strong}
\eeq
Several such predictions are represented by the gray dashed lines in figure \ref{numplot}.

The claim of the current paper is that these numbers can be matched with the weak coupling structure functions.

We will first check this claim by comparing \eqref{intsigma} with the exact tree-level results of \cite{paper1}. The data required to compute $C_{123}$ from the results of that paper are the positions of the Bethe roots of the three operators. For a folded string the Bethe roots are distributed along two symmetric cuts, see figure \ref{foldedcuts}.

Consider first the operator $\O_2$. The total number of Bethe roots is the number of $\bar X$ fields which is equal to $J_1-2$. In each cut we have $(J_1-2)/2$ roots. For $\O_1$ we have $J_1$ scalars $ X$. Hence, operator $\O_1$ is parametrized by $J_1$ Bethe roots, two more roots than $\O_2$. We want to consider an operator $\O_1$ which is \textit{roughly} the complex conjugate of $\O_2$. So, where do we put the extra two roots of $\mathcal{O}_1$?

\begin{figure}[t]
\centering
\def\svgwidth{14cm}
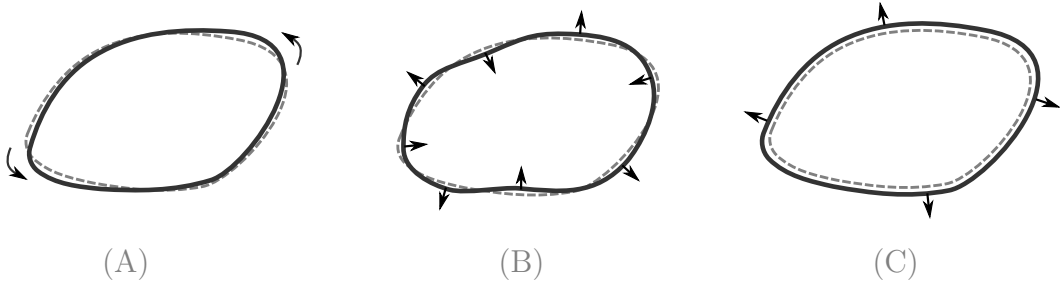
\caption{Three different types of deformations of a classical state:
(A) A global symmetry transformation. (B) A quantum fluctuation.
(C) A fluctuation in an already existing cut (i.e.\ adding a ``zero mode").}\label{backreactions}
\end{figure}

Let us recall what the three natural options are when it comes to adding extra roots to a classical configuration (these are depicted in figure \ref{backreactions}).
\begin{enumerate}[(A)]
\item One option is to put extra roots at infinity. Roots at infinity correspond to acting with a global symmetry generator on the classical configuration, see e.g. \cite{Faddeev:1996iy,BS}. In this case, putting the two roots at infinity would amount to considering the $\O_1$ spin chain state to be obtained from the $\bar\O_2$ state by acting twice with the lowering operator $S^-$. This is the global
symmetry operator which converts a $Z$ field into an $X$.
\item Another option is to add roots at finite values outside the classical cuts. Putting roots outside the cuts at finite values can be interpreted as considering quantum fluctuations around the classical solution \cite{papers}.
\item The last option is to add more roots to the classical cuts which already exist. The number of roots on each of these classical cuts corresponds to  the action variables of the classical solution \cite{vicedo}. Hence adding roots to already open cuts leads to a new classical state with slightly larger magnitude for the same excited action variables.
\end{enumerate}
Naively it might seem like it doesn't matter where we add the extra roots and all of these options should correspond to $\O_1 \simeq \bar \O_2$. This is \textit{not} the case. As will be explained later, there is a sense in which only the last option corresponds to  $\O_1 \simeq \bar \O_2$. So, let us consider that case and add one root to each of the cuts of $\mathcal{O}_2$, which has $(J_1-2)/2$ roots per cut.

We will now provide a numerical check of the agreement between the weak and strong coupling results in the Frolov-Tseytlin limit. We will discuss the general analytic match for general BPS states and general classical solutions later. For the numerics we need to find the solutions to Bethe equations with two symmetric cuts for both $\O_1$ and $\O_2$. This is very simple to do even for a very large
number of roots
(see for example \cite{Bargheer:2008kj}).
 The obstacle is that we then need to plug these solutions into the expressions of \cite{paper1} which involve sums over partitions of Bethe roots, computations of sub-determinants and so on. In practice this means that the  doable numbers of roots is not very large. This is not a serious problem: what we do is compute the three-point function for as many roots as we can and fit the data to estimate the large-length asymptotics.

For example, for $\alpha=1/4$ we find
\begin{center}
\verb'data={{8, 1.27416239226}, {16, 0.976868082293}, {24, 0.919415218869},' \\
\verb'{32, 0.897060720354}, {40, 0.885697954062}, {48, 0.878999431990},'\\
\verb'{56, 0.874655453369}}'
\end{center}
where the first entry is the length of operator $\O_1$ and the second entry is $r=|C_{123}^{\bullet\bullet\circ}/C_{123}^{\circ\circ\circ}|$ computed from \cite{paper1}, see also appendix \ref{reviewap}. Performing a fit as
\begin{center}
\verb'Fit[data,J/J^Range[7],J]'
\end{center}
we find the numerical prediction\footnote{We can also use the numerical data to predict the $1/J$ finite size corrections. For example, for filling fraction $\alpha=1/4$ we find $r=0.856889+ 0.565177/J+\dots$. In principle the first $1/J$ correction should be matched with the (Frolov-Tseytlin expansion of the) one loop computation at strong coupling. It would be extremely interesting to try this.}
\beq
r_{\text{num}} \simeq 0.856889 
\eeq
which agrees neatly with the strong coupling result \eqref{1/3strong} within the numerical error. In figure \ref{numplot} we plotted several other fits and data for several different filling fractions. We see that, very non-trivially, the weak coupling results approach all the predictions from strong coupling in the Frolov-Tsetytlin limit!

\begin{table}
\beq\nn
\begin{array}{c|lll}
\alpha & \text{Prediction}& \text{Extrapolation}&\text{Error} \\
\hline
 \frac{1}{3} & {\bf0.799}4912 & {\bf0.799}2237 & 3\times 10^{-4} \\
 \frac{1}{4} & {\bf0.8568}976 & {\bf0.8568}895 & 8\times 10^{-6} \\
 \frac{1}{5} & {\bf0.88875}95 & {\bf0.88876}01 & 6\times 10^{-7} \\
 \frac{1}{6} & {\bf0.90901}55 & {\bf0.90901}67 & 1\times 10^{-6} \\
\end{array}
\eeq
\caption{Numerical data compared with the analytic prediction for different values of the filling fraction $\alpha$ of operator $\O_1$.}
\la{tab:dt}
\end{table}

In appendix \ref{ex2} we provide another simple example, still using the folded string as our toy model but using a different BPS operator $\O_3=\Tr(Z^j)$ which has been recently studied abundantly in the literature \cite{recentpapers1, recentpapers2, recentpapers3,Roiban:2010fe, common,Buchbinder:2010ek}.

A few words of comfort to the readers less familiar with numerical
fits. Throughout the paper we will perform several fits in order to
extrapolate our finite $J$ data and find the large $J$ asymptotics. We
are given a set of data $\{J , r(J) \}$ with $n$ points and perform a
fit of this data with $n-1$ coefficients, e.g. something like $r=a_0
+a_1/J +\dots + a_{n-1}/J^{n-1}$. We use the data to fix the constants
$a_k$ using Mathematica. Then we plot the resulting fit for values of
$J$ which are typically much larger that the $J$'s in the data (that
is the whole idea of using fits!). The point to emphasize is that the
only assumption we need to make is on the form of the expansion
of $r(J)$. Otherwise \textit{there is no fine-tuning whatsoever}
involved.\footnote{For example, the $\alpha=1/3$ curve in figure
\ref{numplot} has quite a nice non-trivial behavior with a minimum and an inflection
point nicely converging from below to the predicted analytical result.
This is the honest outcome of a fit like the one described in the
text, no fine-tuning is involved. Similarly, in appendix \ref{ex2} we
see that the fits predict nice constant asymptotics for $r(J)$ which
are approached well beyond the range of the data, figure \ref{NumpD}. The match with the
analytics is again perfect without any fine-tuning whatsoever.}

\subsection{Landau-Lifshitz coherent states}
In this section we shall study the classical limit of  $SU(3)$ spin chains, see e.g. \cite{Stefanski:2004cw}. At each site we have three degrees of freedom which we denote as
$
|X\> \, , \, |Y\> \, \text{and} \, |Z\> \,.
$
We use the notation
$
|{\bf u}\> = u_1 |X\>+u_2 |Y\> + u_3 |Z\>
$
to denote a linear superposition of these three degrees of freedom. We normalize ${\bf u}$ to be a unitary vector so that automatically
\beq
\<\bu |\bu\>= \bbu \cdot \bu = 1 \, .
\eeq
 The Hamiltonian we consider is the one arising in the one-loop computation of the Dilatation operator in $\mathcal{N}=4$ SYM. It reads
\beq
H = \frac{\lambda}{8\pi^2}\sum_{n=1}^J \( \mathbb{I}_{n,n+1}- \mathbb{P}_{n,n+1}\)
\eeq
where $\mathbb{P}_{n,n+1}$ is the permutation operator which swaps the degrees of freedom in sites $n$ and $n+1$ while $\mathbb{I}_{n,n+1}$ is the identity. Generic eigenstates are Bethe eigenstates which generalize (\ref{operators}). They are very entangled complex quantum states. A huge simplification occurs in the large $J$ classical limit mentioned in the introduction. In this limit the exact states can be approximated by coherent classical states which simply read \cite{LL}
\beq
|\varphi\> = |{\bf u}^{(1)}\> \otimes \dots \otimes |{\bf u}^{(J)} \> \,.
\eeq
Note that the norm of such states is automatically equal to $1$. In the operator language this state translates into (\ref{cohop}).

The physical picture is that of a slow varying spin wave pointing in direction ${\bf u}$ which slowly changes as we go from one site to the next. It is then convenient to package all the information about the classical state in a continuous field
\beq
\bu(\sigma)= \bu^{(n)} \qquad \text{with} \qquad \sigma = 2\pi\,\frac{n}{J}  \,.
\eeq
To get some experience with these classical states let us review the computation of the energy of such low wave length states
\beqa\la{Energy}
\< \varphi| H | \varphi \> &=&  \frac{\lambda}{8\pi^2 } \sum_{n=1}^{J} \< \bu_{n} | \otimes \< \bu_{n+1}|  \( \mathbb{I}_{n,n+1}- \mathbb{P}_{n,n+1}\) | \bu_{n} \> \otimes | \bu_{n+1} \> \nn \\
 &=& \frac{\lambda}{8\pi^2 } \sum_{n=1}^{J} \[1 - (\bbu_{n}\cdot \bu_{n+1})( \bbu_{n+1}\cdot \bu_{n})\]  \\
 &\simeq& \frac{\lambda}{2 J } \int_{0}^{2\pi} \frac{d\sigma}{2\pi} \, \d_{\sigma}\bar {\bf u}\cdot\d_\sigma {\bf u}\;. \nn
\eeqa
This expression precisely agrees with the strong coupling result (\ref{resultstrong}). By identifying $H|\varphi\>=i\d_\tau|\varphi\>$ in the left hand side of (\ref{Energy}), we reproduce the integrated strong coupling equation of motion (\ref{LLeom}). More generally, the coherent states evolve in time as follows from the Schr\"{o}dinger equation. That evolution is equivalent to the equation of motion (\ref{LLeom}) \cite{LL}.

The Hamiltonian is just one of the many conserved charges of the integrable spin chain, which is a classical operator in the classical limit. The (trace of the) transfer matrix encodes all of them. Hence, a nice way of identifying whether a coherent state mimics an exact state is by measuring the expectation value of the transfer matrix and imposing that it matches the classical limit of the quantum transfer matrix \cite{KMMZ}.

A related but more technical comment is the following. The exact Bethe eigenstates are highest weights, $S^+ |\psi\>=0$. The coherent states which we use to mimic them are only highest weights at the level of averages, that is $\<\varphi | S^+ | \varphi\>=0$. Similarly, the exact states are cyclic, $e^{iP} |\psi\>=|\psi\>$ while  $\<\varphi | e^{i P} |\varphi\>=1$. Cyclicity of the exact states means that we don't need to care about where we break them and so on when computing expectation values or three-point functions. On the other hand, to mimic these computations with coherent states we should align the operators as done in the following section.

\subsection{Heavy-Heavy-Light three-point functions} \la{coherentderivation}
In this section we present the idea of the derivation of the general match between weak and strong coupling using the coherent state machinery reviewed in the previous section. A rigorous proof would involve cluttering the text with lots of kets, bras, tensor products etc without great benefit of clarity. Instead we sketch the derivation and the main physical ideas, leaving to the diligent reader the pleasure of filling in the details.

One of the main issues in a rigorous derivation is what we mean by $\O_2$ been \textit{roughly} the complex conjugate of $\O_1$.
Naively one might think that the precise details of how $\O_2$ differs from $\bar \O_1$ should not matter. However we will show that it does and hence the complication, see sections \ref{ex3} and \ref{genback}. In any case, since we will be more precise about this point later, we will sacrifice rigor in this section.

Having the necessary machinery, we move to the main object of study of this paper. We will consider the computation of three-point functions where two of the operators $\O_1$ and $\O_2$ are classical operators (well approximated by coherent operators/states) while the third operator $\O_3$ is a small operator. To simplify the expressions we will use the notation $\mathcal{L}_i = (\text{length of operator $\O_i$})/2\pi$.

For now, we take $\O_3$ to be a vacuum descendant,
\beq
\O_3={\cal N}\({\rm Tr}\[\bar X^{j_1} \bar Y^{j_2} Z^{j_3}\]+\text{permutations}\nn\) ,
\eeq
where the normalization coefficient ${\cal N}$ is simply one over the square root of the number of permutations.
We only distinguish permutations up to a cyclic permutation due to the cyclicity of the trace
\beq
{\cal N}=\sqrt{\frac{j_1!j_2!j_3!}{(j-1)!}}\;.
\eeq
To accomplish our businesses with combinatorics let us notice that among
all the terms in $\O_3$ only the ones of the form ${\rm Tr}[Z^{j_3}\,\text{anything}]$ can give a nonzero contribution
to the Wick contractions with $\O_1$ and $\O_2$, see figure \ref{HHL}b. The number of such terms is obviously
 equal to $\binom{j_1+j_2}{j_1}$.

The operators $\O_1$ and $\O_2$ are described by coherent states as introduced in the previous section. In the operator language they are given by (\ref{cohop}), while in the spin language we have
\beqa
| \O_1 \>&=& \dots \otimes\left|{\bu}(\tfrac{n}{\mathcal{L}_1}) \right\> \otimes\left| {\bu}(\tfrac{n+1}{\mathcal{L}_1}) \right\>\otimes \dots \\
\< \O_2 |&=& \dots \otimes\left\<{\bbu}(\tfrac{n}{\mathcal{L}_2}) \right|\otimes \left\< {\bbu}(\tfrac{n+1}{\mathcal{L}_2}) \right| \otimes\dots
\eeqa
The three-point function is now obtained by Wick contracting the three operators as depicted in figure \ref{HHL}b. Since there is no entanglement in operators $\O_1$ and $\O_2$ the contractions are trivial. The operator $\O_3$ is glued, that is Wick contracted, with the other states at sites $k,k-1,...\,$. We should then sum over the insertion starting point $k$. The Wick contractions between $\O_1$ and $\O_2$ give
\beq
{\cal I}_k=\prod_{i=k+1}^{L+k}\bbu(i/\mathcal{L}_2)\cdot\bu(i/\mathcal{L}_1) \,.\la{other}
\eeq
where $L$ is the number of contractions between $\O_1$ and $\O_2$. What is left is to consider the Wick contractions between $\O_3$ and $\O_1$, $\O_2$.
Consider one of the terms in $\O_3$ which gives a nonzero contribution ${\rm Tr}[Z^{j_3}\bar X\bar X\bar Y\dots]$.
The Wick contraction of this operator with the coherent states $\O_1$, $\O_2$ at position $k,k-1,\dots\,$ gives
$$
\,\bar u_3(\tfrac{k}{\mathcal{L}_2})\bar u_3(\tfrac{k-1}{\mathcal{L}_2})
\dots
\bar u_3(\tfrac{k-j_3+1}{\mathcal{L}_2})
u_1(\tfrac{k}{\mathcal{L}_1})
u_1(\tfrac{k-1}{\mathcal{L}_1})
u_2(\tfrac{k-2}{\mathcal{L}_1})\dots \;.
$$
Since the field changes slowly over those sites, we can approximate this contribution by
\beq
\,u_1^{j_1}(k/\mathcal{L}_1)\, u_2^{j_2}(k/\mathcal{L}_1)\,\bar u_3^{j_3}(k/\mathcal{L}_2)\nn\;.
\eeq
Thus, all nonzero Wick contractions between
$\O_3$ and $\O_1$, $\O_2$ give approximately the same contribution
and there are $\binom{j_1+j_2}{j_1}$ of them. Hence, the result of these contractions is
\beq
{\cal J}_k=\mathcal{N}\binom{j_1+j_2}{j_1}\,u_1^{j_1}(k/\mathcal{L}_1)\, u_2^{j_2}(k/\mathcal{L}_1)\,\bar u_3^{j_3}(k/\mathcal{L}_2)\nn\;.
\eeq
All together, we get
\beq\la{contract}
C_{123}\simeq \,\sum_{k=0}^L{\cal I}_k\,{\cal J}_k\,\,,
\eeq

Note that we have a $U(1)$ gauge invariance when
representing the coherent states as tensor products of single site states.
That is, we can multiply the state associated with any site of the chain by an independent phase factor
\beq\la{gauge}
\bu(j/\cL)\to e^{i\Lambda(j)}\,\bu(j/\cL)\ ,\qquad\bbu(j/\cL)\to e^{-i\Lambda(j)}\,\bbu(j/\cL)\;.
\eeq
All observables, including the structure constant (\ref{contract})
are of course invariant under the gauge transformation
however ${\cal I}_k$ and ${\cal J}_k$ in (\ref{contract}) do transform nontrivially
 (for $j_3\ne j_1+j_2$).
We can profit from this fact to choose the most convenient gauge.
Namely we can fix a conformal-like gauge by demanding that ${\cal I}_k$ does not depend on $k$. That is
\beq
1=\frac{{\cal I}_{k}}{{\cal I}_{k-1}}={\bbu(k/\cL_2+L/\cL_2)\cdot\bu(k/\cL_1+L/\cL_1)\over \bbu(k/\cL_2)\cdot\bu(k/\cL_1)}\nn \,.
\eeq
We have
\beqa
\nn \bu(k/\cL_1+L/\cL_1)&\simeq&\bu(k/\cL_1)\ -\ \frac{j_1+j_2}{\cL_1}\d_\sigma\bu(k/\cL_1)\ +\ \O(1/\cL_1^2)
\eeqa
and similarly for ${\bf \bar u}$. Thus we have to require that
\beq\la{Virasoro}
(j_1+j_2-j_3)\ {\bbu}\cdot\d_\sigma\bu=0\;,
\eeq
which matches nicely the Virasoro constraint (\ref{virasoro}). In the conformal-like gauge (\ref{Virasoro}), we further have that
\beq
{\cal I}_0=\prod_{i=1}^L\bbu(i/\cL_2)\cdot\bu(i/\cL_1)\simeq \exp \int\limits_0^{2\pi-{j_3\over\cL_2}}
J  \frac{d\sigma}{2\pi}\,\log\(\bbu\cdot\[1-{j_1+j_2-j_3\over\cL}\sigma\d_\sigma\]\bu\)=1+O(1/\cL)\nn
\eeq
We conclude that in this gauge:
\beq
C_{123}^{\bullet\bullet\circ}=
\frac{(j_1+j_2)!}{j_1! j_2!}\sqrt{j\frac{j_1! j_2!j_3!}{j!}}\int\limits_0^{2\pi}
J\frac{d\sigma}{2\pi}\, u_1^{j_1}\, u_2^{j_2}\,\bar u_3^{j_3}\ ,
\la{mainweak}
\eeq
which is precisely the strong coupling result (\ref{mainstrong})! This is our main result.
Notice also that if we substitute $v_a$ from \eq{BPS} instead of $u_a$ we get
\beq
C_{123}^{\circ\circ\circ}=
J\frac{(j_1+j_2)!}{j_1! j_2!}\sqrt{j\frac{j_1! j_2!j_3!}{j!}}\, \(\frac{J_1}{J}\)^{j_1/2}
\(\frac{J_2}{J}\)^{j_2/2}
\(\frac{J_3}{J}\)^{j_3/2}
\eeq
which is indeed the limit $J_a\gg j_a$ of the known BPS formula \eq{su3bps}.

\subsection{Non-BPS operator $\O_3$}\la{nonBPS}
Our tree-level computation did not rely in any crucial way on the small operator being protected.\footnote{Three-point functions of two heavy operators and certain light non-BPS operators were discussed in \cite{Roiban:2010fe}.} We can therefore trivially generalize it to any small $SU(3)$ operator $\O_3$ as we will now explain. All the manipulations of the previous subsection involving the large operators were not sensitive to the precise form of the small operator and only depended on its global charges.
The only difference when considering non-protected small operators is in the combinatorial factor outside the integral in (\ref{mainweak}). That is, the more general weak coupling result is
\beq\la{nonBPSC}
C_{123}^{\bullet\bullet\bullet}={\cal D}\int\limits_0^{2\pi}
J\frac{d\sigma}{2\pi}\, u_1^{j_1}\, u_2^{j_2}\,\bar u_3^{j_3}\ ,
\eeq
where ${\cal D}$ is the sum of the coefficients of the traces in $\O_3$ in which all the $Z$ fields are next to each other.

To illustrate this point let us consider a simple example where $\O_3$ is the Konishi operator
\beq
\O_3={1\over\sqrt 3}\Bigl[\Tr\(Z^2\bar X^2\)-\Tr\(Z\bar XZ\bar X\)\Bigr]\quad\Rightarrow\quad{\cal D}={1\over\sqrt3} \quad\Rightarrow\quad {{C_{123}^{\bullet\bullet\bullet}}\over C_{123}^{\bullet\bullet\circ}} ={1\over\sqrt 3}\sqrt{3\over2} \, .
\eeq
This ratio is exact and does not rely on the classical limit. Note that although in this example $\O_3$ is in the $SU(2)$ sub-sector the operators $\O_1$ and $\O_2$ can be generic $SU(3)$ operators. Another simple example is
\beq
\O_3=\frac{1}{\sqrt{2}}\Big[ \Tr (Z \bar X \bar Y) - \Tr (Z \bar Y \bar X) \Big] \qquad \Rightarrow \qquad \mathcal{D}=0\,.
\eeq
If we  restrict to the $SU(2)$ sector considered in \cite{paper1}, see figure \ref{3ptfunction}, we have\footnote{For $\O_3$ in the $SU(3)$ case, the first equal sign in  (\ref{simpratio}) still holds provided we take the classical limit.
}
\beq\la{simpratio}
{{C_{123}^{\bullet\bullet\bullet}}\over C_{123}^{\bullet\bullet\circ}}={{C_{123}^{\circ\circ\bullet}}\over C_{123}^{\circ\circ\circ}}={\caA(j_1|\{{w}\})\over\caB(\{{w}\})} \, ,
\eeq
where we have used that the dependence on $\O_1$ and $\O_2$ has canceled out and $\caA$, $\caB$ were introduced in \cite{paper1}. Recall that  we are now considering non-BPS operators $\O_3$ which can be parametrized by its Bethe roots; those are the $w_j$. Again, this formula is exact and does not rely on any classical limit.

For non-protected operators however, we don't expect a match with strong coupling in the classical limit considered so far. That can be seen already at one loop. A main ingredient for the match between strong and weak coupling in the Frolov-Tseytlin limit is that the perturbative expansion organized itself in powers of $\lambda/J^2=1/\kappa^2$. At one loop, there are two types of corrections \cite{Okuyama:2004bd}. One is by insertion of the one-loop Hamiltonian between the operators (see figure \ref{One_loop}). The other is the two-loop correction to the Bethe wave functions. In contradistinction to the case where the small operator is BPS,  these two types of corrections are not suppressed by $1/J^2$ and therefore, do not follow the Frolov-Tseytlin scaling.

If, on the other hand, we take the double scaling limit $1\ll j\ll J$ with $\lambda/j^2$ small then  the operator $\O_3$
 can \textit{also} be treated classically while its back-reaction on the large operators can still be neglected. In such limit, the weak and strong coupling structure constants might match. In the classical limit, we have computed the ratio $\caA/\caB$ (\ref{simpratio}) in the past by simplifying the exact expression and reported the result in \cite{paper1}. Combined with (\ref{simpratio}), we get\footnote{Up to a pre-exponent. That is, $\Gamma={O}(1)$ and this formula does not capture $1/j$ finite-size corrections to $\Gamma$. In particular, the $\sigma$ integral in $\Gamma$ can be replaced by a its saddle point value.}
\beq
C_{123}^{\bullet\bullet\bullet}= \exp\( \,j \Gamma+\O(j^0) \) \la{cool}
\eeq
where
\beq
\Gamma = \frac{1}{j} \log\[\int\limits_0^{2\pi}
J\frac{d\sigma}{2\pi}\, u_1^{j_1}\,\bar u_3^{j_3}\]+\frac{1}{j}\int\limits_0^1 dt \(\,\oint\limits_{\cup \,\mathcal{C}_k} \frac{d\mathtt u}{2\pi i}  \,q\log (2\sin (t q/2))-   \int\limits_{\cup \,\mathcal{C}_k} d{\mathtt u} \,\rho\,\log\(2\sinh(\pi t \rho)\)\,\) \,.\nn
\eeq
Let us summarize the meaning of the several symbols in this formula. The operator $\O_2$ is a classical operator that can be approximated by an $SU(2)$ coherent state parametrized by the Landau-Lifshitz field $u_1(\sigma)$ and $\bar u_3(\sigma)$ as reviewed in the previous sections. The operator $\O_1$ is \textit{roughly} the complex conjugate of $\O_2$ in the sense (C) discussed in section \ref{Weak_coupling}. The operator $\O_3$ contains  $j_1$ scalars $\bar X$, $j_3$ scalars $Z$ and $j=j_1+j_3$. The $j_1\gg1 $ Bethe roots of this operator organize themselves into several cuts $\mathcal{C}_k$. The density of roots in those cuts is $\rho(\mathtt v)$. Finally,
\beq
q({\mathtt u})\equiv \frac{j_1}{\mathtt u}-\int\limits_{\cup \,\mathcal{C}_k} d{\mathtt v} \,\frac{\rho({\mathtt v})}{\mathtt u-\mathtt v} \la{qdef}
\eeq
is a sort of trimmed quasi-momenta. For more details, we refer the reader to \cite{KMMZ,paper1,In_progress}. It would be very nice if that result could be reproduced at strong coupling. A first place to look at might be the small spike solution considered in \cite{recentpapers1}.

\subsection{Back-reaction numerically} \la{ex3}
In this section we begin our study of back-reaction. As before, we start by some numerics to gain intuition about the general picture. We will again use the folded string introduced in section \ref{ex1} as our toy model.
A general analytical discussion is presented in the next section.

As  explained in the introduction the states or the eigenfunctions of the dilatation operator
at one loop are labeled by the set of Bethe roots $u_j=\frac{1}{2}\cot\frac{p_j}{2}$.
Then the states corresponding to a classical finite-gap solution
correspond to configurations of Bethe roots like the one shown in figure \ref{foldedcuts},
where the roots are forming some dense cuts in the complex plane.

The state $\O_2$ in general should have a different number of Bethe roots than the state $\O_1$
since the number of roots is related to the R-charge. Then, as  discussed in the section \ref{ex1}, there are three major possibilities as to
where the extra roots can go:
\begin{enumerate}[(A)]
\item One option is to put extra roots at infinity (i.e.\ with zero momenta $p_j$). Roots at infinity correspond to acting with a global symmetry generator on the classical configuration, see e.g. \cite{Faddeev:1996iy,BS}.
\item Another option is to add roots at finite values outside the classical cuts.
\item The last option is to add more roots to the classical cuts which exist already.
\end{enumerate}

In section \ref{ex1} the last possibility was examined numerically and a perfect match with the
analytical result (\ref{mainweak}) was found. In this
section we discuss the numerical results concerning the first two options.

\subsubsection{Numerical results for roots at infinity}
\begin{figure}[t]
\centering
\def\svgwidth{13cm}
\includegraphics{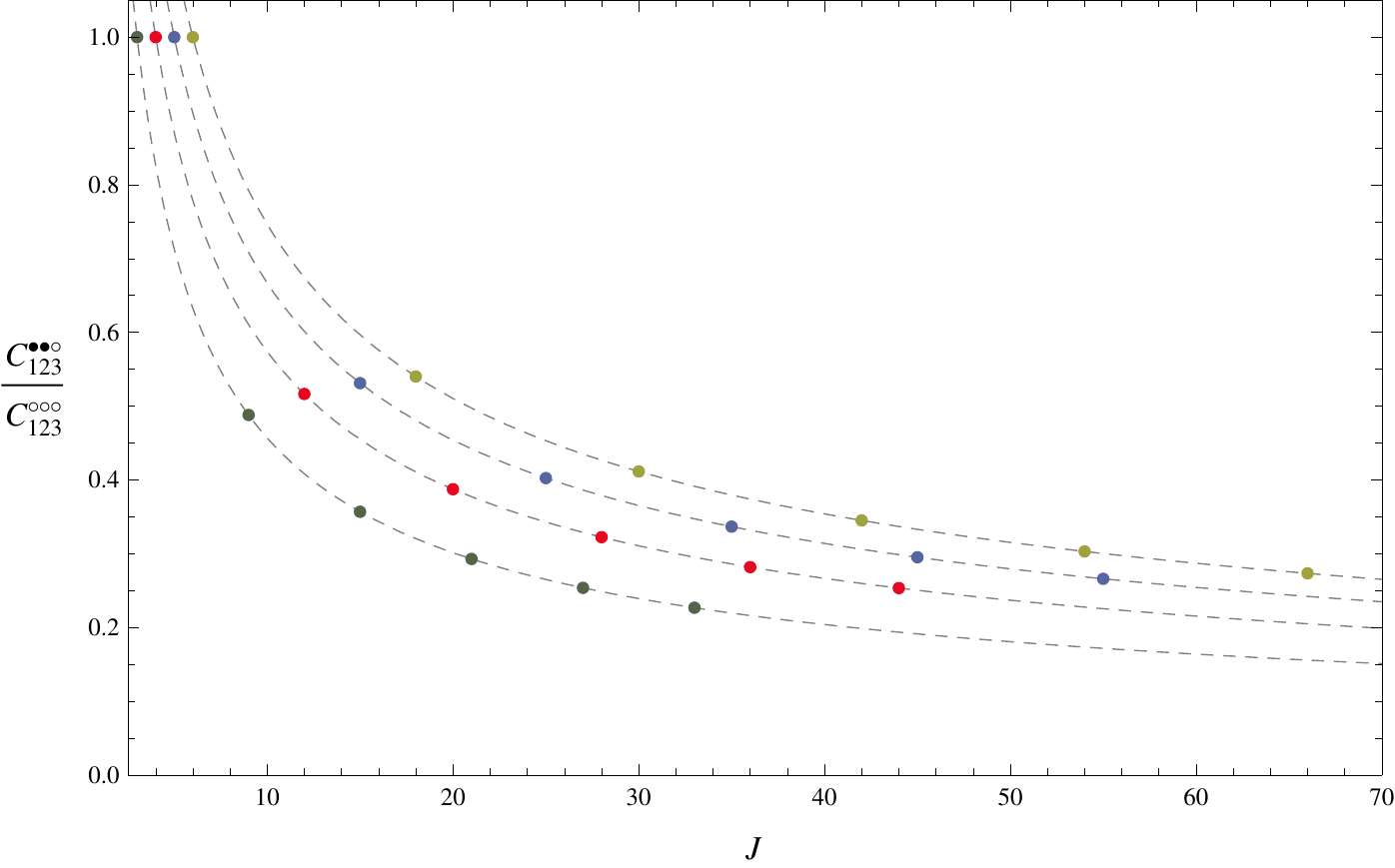}
\caption{Numerical data for the folded string example for the case (A) against the analytical prediction (dashed gray lines)
for different values of the filling fraction of operator $\O_1$: $\alpha=1/3,1/4,1/5,1/6$, from bottom to top.
}\label{numplot2}
\end{figure}

Here we study the option (A) from the list above.
We consider $\O_3$ to be a BPS operator with $j/2$ scalars $\bar X$ and $j/2$ scalars $Z$ for even $j$.
In this case the final operator $\O_1$ and $\O_2$ have the same length, but $\O_1$ has $n\equiv j/2$ more Bethe roots than the operator $\O_2$.
We consider the situation when the extra $n$ roots are sent to infinity,
which means that $\O_1$ is a descendant of $\bar\O_2$.

For the simplest case $n=1$ we can compute the result exactly. In appendix \ref{O3S-} we show that
\beq
\frac{C_{123}^{\bullet\bullet\circ}}{C_{123}^{\circ\circ\circ}}=\frac{1}{\sqrt{J}}\sqrt{\frac{1-2\alpha+2/J}{\alpha(1-\alpha+1/J)}}\;.
\eeq
This suggests that in general one should have
\beq\la{infroots}
\frac{C_{123}^{\bullet\bullet\circ}}{C_{123}^{\circ\circ\circ}} \sim \frac{1}{J^{n/2}}\;.
\eeq
Indeed the numerics for the case $n=2$ and $\alpha=1/5$ give
\beq\nn
\begin{array}{c | c c c c c c}
J& 10 & 20 & 30 & 40 & 50 & 60 \\                                \hline
C^{\bullet\bullet\circ}_{123}/C^{\circ\circ\circ}_{123} & 1 & 0.27036118 & 0.14195150 & 0.092991262 & 0.068076848 & 0.053258866
\end{array}
\eeq
Fitting the
data by $J^a\sum_{i=0}^4\frac{b_i}{J^i}$ we obtain $a=-1.01$, which is indeed consistent with \eq{infroots}.\footnote{We also considered cases with more Bethe roots at infinity and found again evidence for (\ref{infroots}) although the precision decreases substantially as we increase the number $n$ of Bethe roots sent to infinity.}
Notice that for exactly the same $\O_2$ and $\O_3$, but in the situation when all roots belong to the big classical cuts of $\O_1$,
we obtained a finite value $r=0.88876$ by fitting the numerical data to infinite length (see table \ref{tab:dt}). Thus we see that back-reaction is very important in this
case and the result is completely different compared to the option (C) studied in section \ref{ex1}.

In the next section we give a simple reason for that behavior and draw
the general criteria for back-reaction.

\subsubsection{Numerical results for finite roots}

To keep the zero momentum condition in $\O_1$, the two extra finite roots that we add should have opposite mode numbers $k$ and $-k$. The case $k=1$ is the one considered in section \ref{ex1}; it amounts to adding more roots to the already open classical cuts. The cases $k=2,3,\dots$ corresponds
to the case (B) from the list above.

\begin{figure}[t]
\centering
\def\svgwidth{14cm}
\includegraphics{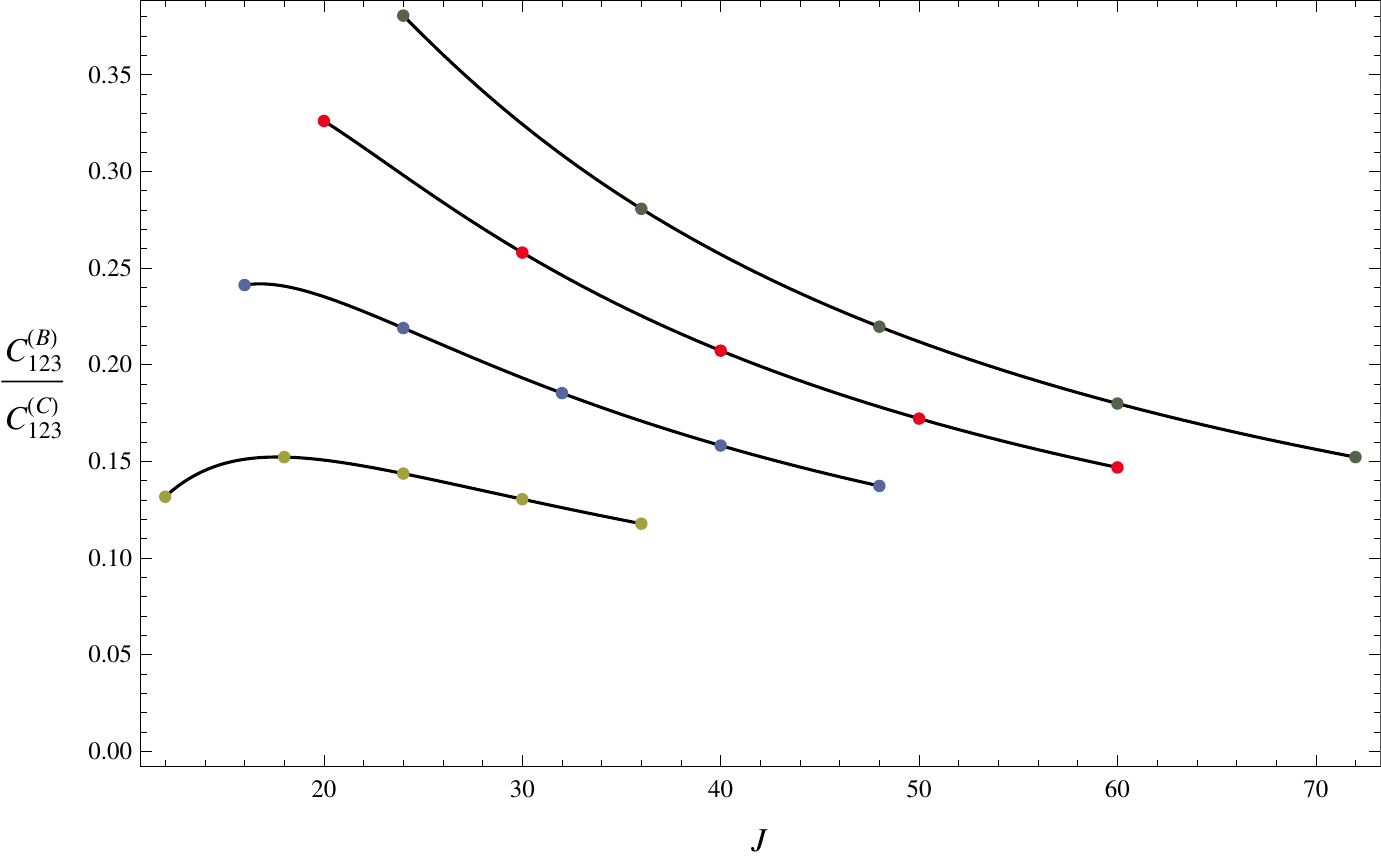}
\caption{Numerical data for the three-point function with operator $\O_1$ deformed by adding
two finite roots with mode numbers $k=\pm3$ (option (B)) divided by the
same three-point function with operator $\O_1$ deformed by adding
two roots to the already existing cuts with mode numbers $k=\pm1$ (option (C))
for different filling  fractions of operator $\O_1$: $\alpha=1/3,1/4,1/5,1/6$ (from bottom to top).
We see that the ratio does not go to one. In other words, back-reaction is relevant.
}\label{numplot3}
\end{figure}

The main question is whether we get the same result when
adding the two extra roots with mode number $k \neq  1$ compared
to the case $k=1$ considered in section \ref{ex1} or not.
To answer this question we use again the analytic results \cite{paper1} and evaluate these expressions numerically. The results are shown in figure \ref{numplot3}.
We conclude that again the answer is no
and that the result differs considerably. In other words, back-reaction is relevant.

\subsection{General criteria for back-reaction} \la{genback}

In this section we will give a qualitative explanation for the dependence of the result on the type of modification of the state $|\O_1\>$ with respect to the state $|\bar\O_2\>$
which is necessary for R-charge conservation.
As we discussed above there are three options (see figure \ref{backreactions}).
The case (A) we mainly discuss in appendix \ref{moreonbackreaction}.
One can think about the case (A) as a limit of (B) when the mode number $k$ of the extra roots goes to zero. In this section we focus therefore on cases (B) versus case (C).

To understand better the difference between these two cases, we first consider the example of a free massless scalar field $\phi(\sigma,\tau)$ on the cylinder. A general classical solution is of the form:\footnote{For simplicity, we only consider the left-moving solutions.}
\beq
\phi(\sigma,\tau)=\sum_n A_n \cos\(n(\sigma+\tau)+\phi_n\)\;.\nn
\eeq
The analog of the finite gap solution would be the situation when the sum has only finitely many terms.
The amplitude $A_n$ is related to the number of excitations inside a cut through a Bohr-Sommerfeld quantization condition, which in the present case coincides of course with the oscillator
$nA_n^2=c N_n$
where $c\sim \hbar$
is a small constant and $N_n$ is the number of excitations at mode number $n$ called {\it filling number}. We see that
\beq\la{quantization}
A_n\simeq \sqrt{N_n\over n}\ .
\eeq

Suppose we start with an unperturbed solution consisting of a single large cut
\beq
\phi_0(\sigma,0)=A \cos(n\sigma)\;.\nn
\eeq
There are two different types of perturbations we can introduce. One is by adding a small number of excitations, $l$, to the existing large cut
\beq
\phi^{(\text{C})}(\sigma,0)=\[A+\delta A^{(\text{C})}\] \cos(n\sigma)\;.\nn
\eeq
The other, is by adding the small number of excitations $l$ at a different mode number $k\neq n$
\beq
\phi^{(\text{B})}(\sigma,0)=A\cos(n\sigma)+\delta A^{(\text{B})}\cos(k\sigma)\;.\nn
\eeq
These perturbations obviously correspond to the options (C) and (B) in the interacting case.
Since the amplitude is proportional to the square root of the filling number (\ref{quantization}), for the case $N\gg l$ we get
\beq\la{behave}
\delta A^{(\text{C})} \simeq A\frac{l}{2N}\;\;,\qquad\delta A^{(\text{B})} \simeq\sqrt{l\over k}\simeq A\sqrt\frac{nl}{Nk} \, .
\eeq
We see that in these cases the scaling of the correction to the wave function is very different.

Having understood this let us turn back to our calculation of the three-point function at weak coupling and see how the difference between the two behaviors in (\ref{behave}) really matters. The only difference between the quasi-classical approximation for the Bethe eigenstates and the free field example is that now the specific form of the solutions are more complicated and when we add new roots, the old roots get a small correction. These differences are irrelevant for the behavior (\ref{behave}).

In the large $J$ limit, the quasi-classical approximation is valid and we can repeat our previous calculation of the three-point function as prescribed in section \ref{coherentderivation}. For simplicity we consider the case $j_3=j_1+j_2$ so that
the lengths ${\cal L}_1={\cal L}_2$ are equal and we do not have to worry about the $U(1)$ gauge transformations (\ref{gauge}). This time however, let us be a bit more careful about back-reaction. Namely we assume that the coherent states for $\O_1$ and $\O_2$ are a bit different and are parametrized
by two different functions $\bu_2(\sigma)\equiv\bbu(\sigma)$ and $\bu_1(\sigma)\equiv\bu(\sigma)+\delta\bu(\sigma)$.
Since $\bu_1$ is a unit vector, we have
\beq\la{normu2}
 \bbu\cdot \delta\bu+ \delta\bbu\cdot {\bf u}= - \delta\bbu \cdot\delta\bu\;.
\eeq
An important step in the previous `naive' calculation was the proof that the direct Wick contractions between $\O_1$ and $\O_2$ (\ref{other}) are trivial. The overlap of the two coherent wave functions is equal to
\beq
{\cal I}\equiv \exp\(\frac{J}{2\pi}\int d\sigma \log (1+\bbu\cdot \delta{\bf u})\)
\simeq
\exp\(\frac{J}{2\pi}\int d\sigma \[\bbu\cdot \delta{\bf u}-\frac{1}{2}(\bbu\cdot \delta{\bf u})^2\]\)
\eeq
As we are only interested in the absolute of the three-point function, we only need to consider the real part of the integrand. The real part of the first term in the integrand is precisely the left-hand side of (\ref{normu2}). Thus we see that the real part of the integrand is of order $\sim J\|\delta {\bf u}\|^2$.

For the type of correction (C) where the roots are added to the big open cuts, we found that $\delta {\bf u}^{(C)}\sim l/J$. In that case, the integrand goes as $l^2/J$ and the overlap
${\cal I}$
is indeed $1$. That lead us to the main result (\ref{mainweak}). For the second type of corrections where the roots are not added to the big cuts, we found that $\delta {\bf u}^{(B)}\sim \sqrt{l/J}$. In that case we then get
\beq\la{corr}
|{\cal I}|\simeq e^{-c\, l}
\eeq
and thus for this case we have an additional  exponential suppression for large $l$ (the coefficient must be negative
since the overlap cannot exceed $1$). Furthermore, in this case, (\ref{corr}) is not even the only correction. This is indeed the case according to what we found from numerics. That is, for small $l$'s we found an order $1$  correction compared to case (C) studied in section \ref{ex1}.

Note that if we try to pass to the limit $k\to 0$ corresponding to the case (A), the wave function correction
goes to infinity as one can easily see from \eq{behave}. This indicates that when adding roots at infinity, the suppression should be even stronger in agreement with our finding (\ref{infroots}).

We end this section by noting that at strong coupling, the classical operators are described by the same coherent states. Our findings in this paper indicate that also at strong coupling, the three-point function of two large and one small operators is dominated by the overlap of the wave functions. As such, we expect the above criteria for back-reaction to apply.

\section{Conclusion and open problems}
In this paper we considered the computation of structure constants $C_{123}$ of single trace gauge-invariant operators in $\mathcal{N}=4$ SYM.

We further focused on a classical limit where the operators $\O_1$ and $\O_2$ are made out of a large number of constituent fundamental fields while $\O_3$ is a protected BPS operator which is taken to be small compared to the other two.
This means that $\O_2$ is a classical operator, $\O_3$ is a quantum operator and $\O_1$ is \textit{roughly} the complex conjugate of $\O_2$.\footnote{We cannot have simply $\O_1=\bar \O_2$ because of R-charge conservation.} Under a proper identification of $\O_1 \simeq \bar \O_2$ we obtained a perfect match between the leading weak coupling results and the strong coupling results in the Frolov-Tseytlin limit, see also figure \ref{HHL2}. 
The match holds for \textit{any} classical string solutions with nontrivial motion in $S^5$ and which admit such limit.
Our result is analogous to the match observed in the spectrum problem \cite{KMMZ,BKSZ} and we hope it can be equally inspiring in guessing an all-loop interpolation for structure constants as done for the spectrum in \cite{AFS,BS}.   

At the same time we also found that the precise form of $\O_1 \simeq \bar \O_2$ \textit{does} matter. We get a match for a particular (albeit natural) choice of $\O_1$. This choice can be described as follows. The classical operator $\O_2$ is described by many Bethe roots which condense into cuts representing a classical wave function \cite{KMMZ}. One such example is depicted in figure \ref{foldedcuts}. Then, in the setup considered in this paper, $\O_1$ corresponds to an operator with some more Bethe roots. The question is then, where do we put them? The answer is quite simple: we should add them to the classical cuts that are already present in the classical operator $\O_2$.\footnote{According to our discussion in section \ref{genback}, we expect that if the classical solution contains many classical cuts we can add the extra roots to any of the existing cuts and get the same classical result.}

The three-point function of two classical operators and one quantum operator is dominated by the overlap of the two classical operators. If instead of adding the extra roots to the open cuts we put them outside, then the change in the classical operator is not small enough. Putting roots outside the cuts at finite values can be interpreted as considering quantum fluctuations around the classical solution \cite{papers}. Putting extra roots at infinity corresponds to acting with global symmetry generators on the classical solution. In both of these cases, the classical state $\O_1$ still seems to be very similar to $\bar O_2$. However, for these cases the structure constants are very sensitive to the precise details of the deformations and do not match the strong coupling result. The match only occurs  when the roots are added to previously open classical cuts.

In addition we computed the structure constant between three non-protected classical operators with one of the operators still much smaller than the other two, see  (\ref{cool}). It would be great if this could be matched with strong coupling. The asymptotic analysis of the small spike from \cite{recentpapers1} might be of great use in this case.

One of the main ingredients of this paper was the use of coherent states for studying the weak coupling classical operators. The other main ingredients were the exact tree-level formulae for the structure constants derived in \cite{paper1} and evaluated numerically in the present paper. They allowed us to be confident about the validity of the coherent state approach and they also provided very convincing evidence for the physical picture vindicated in the  previous paragraphs.

For example, for the folded string we solved the Bethe equations numerically and plugged the solution into our analytic results \cite{paper1}. We then compared the answer with the analytical predictions from coherent states and found a precise agreement in the classical limit.

Note that the exact expression for the type of structure constants considered here is very complicated \cite{paper1}. Here, in the classical limit, we have obtained a huge simplification of it, see (\ref{mainweak}). It would be very interesting to understand how to derive the simplified result directly from the exact expression. Such understanding may teach us how to do the same for the case where all three operators are generical classical operators.\footnote{Note that the exact expressions for the two cases are written in terms of the same building blocks. In a similar case where only one of the three operators is classical whereas the other two are BPS, we have already succeed in simplifying the exact expression. The result in that case is an exponent of a non-trivial functional of the classical data \cite{paper1}.}

For the computation of the classical three-point functions, all we really needed was the fact that the large operators are well approximated by coherent states. Such semiclassical coherent state approximation exists for any spin chain in the appropriate limit and hence our methods can be used for any large $N$ gauge theory. No need for integrability (the match with strong coupling is of course very special to ${\cal N}=4$ SYM).

There are several interesting possible extensions of this work. One of these is to extend the analysis done in this paper  for the $SU(3)$ sector into other sectors as well as the full theory.
At weak coupling, even for the full $SU(3)$ sector the exact expression is not yet known. However, as done in this paper, it might be  easier to derive the classical results directly in the coherent states language rather than to simplify the exact expressions.

\begin{figure}
\centering
\def\svgwidth{6cm}

\ifpdf
    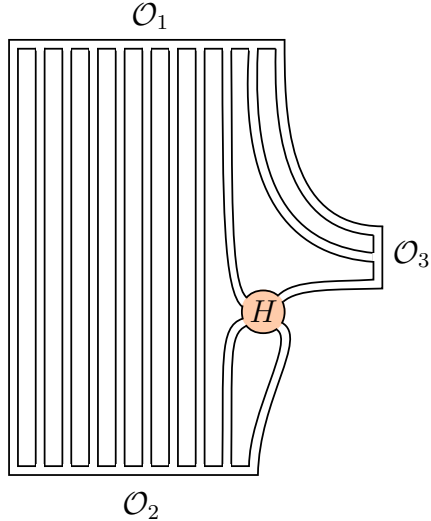
\else
    \com{PDF-picture replacement}
\fi

\caption{At one loop, the three-point function gets two types of corrections. One type is the two-loop correction to the external wave functions eigenstates. The other correction shown in this figure comes from the one-loop Hamiltonian insertion $H$ between legs that originate from one of the operators and goes to two different operators. As opposed to the overlap of the wave functions, that contribution is very local and therefore it is somehow simpler. It can be thought of as measuring the energy cost of splitting the string \cite{Okuyama:2004bd}. Outside the scalar sector the picture seems to be more involved \cite{Alday:2005nd}.}\label{One_loop}
\end{figure}

Another important direction would be to extend this work to one loop. It is simple to expand the strong coupling result (\ref{gen}) to next order in $1/\kappa^2$. At weak coupling, one should decorate the tree-level contractions by inserting the one-loop dilatation operator \cite{Okuyama:2004bd,Roiban:2004va,Alday:2005nd} as represented in figure \ref{One_loop} and correct the wave functions. An optimistic scenario would be an agreement up to two loops.

Finally, it would be interesting to extend our weak coupling results to higher-point functions. In some specific limits, we expect the same techniques to apply. One such case is the four-point function of four large operators that are designed such that their four-point function is dominated by the exchange of a finite set of small operators. Another case is the four-point function of two large classical operators and two small BPS operators \cite{Roiban:2010fe,Buchbinder:2010ek}.
In this case we expect that only a simple modification of our result is needed.

Of course, the most exciting future direction would be to address the question of the discretization of the classical results and then conjecture the all-loop result for the structure constants, at least for operators that are asymptotically large. In the past, for the spectrum problem, this line of attack was extremelly fruitful, see e.g. \cite{AFS,BS,Beisert:2010jr}.

\section*{Acknowledgments}

We thank J.\ Caetano, N.\ Drukker, R.\ Janik, J.\ Maldacena, T.\ McLoughlin, J.\ Penedones, S.\ Schafer-Nameki, A.\ Tseytlin, K.\ Zarembo for many interesting discussions. We also thank J.\ Maldacena and K.\ Zarembo for very useful comments on the manuscript. The research of J.E., A.S. and P.V. has been supported in part by the Province of Ontario through ERA grant ER 06-02-293. Research at the Perimeter Institute is supported in part by the Government of Canada through NSERC and by the Province of Ontario through MRI. A.S. and
P.V. would like to thank KITP for warm hospitality during the period of completion of this work.
N.G. would like to thank Perimeter Institute for hospitality during the embryonic stages of this work.
This work was partially funded by the research grants PTDC/FIS/099293/2008 and CERN/FP/109306/2009. This research was supported in part by the National Science Foundation under Grant No. NSF PHY05-51164.

\appendix

\section{Tailoring numerics}\la{reviewap}
According to the results of \cite{paper1}, the ratio $C_{123}^{\bullet\bullet\circ}/C_{123}^{\circ\circ\circ}$ can be computed in Mathematica as
\beq
r={\verb"Cxxo[L1,N1,L2,N2,L3,N3,us,vs]" \over \verb"Cooo[L1,N1,L2,N2,L3,N3]"} \, ,
\eeq
where \verb"Cxxo" and \verb"Cooo" are defined in appendix D of \cite{paper1} and \verb"us" and \verb"vs" refer to the roots of operators $\O_1$ and $\O_2$, respectively.\footnote{For simplicity, the discussion in this appendix refers to the case when all roots in the heavy operators are finite. After some slight modifications described in the main text of \cite{paper1}, the codes given in the appendices of \cite{paper1} can (and were) used to compute the cases when some roots are at infinity.} For example, if we wanted to reproduce the first point for $\alpha=1/4$ in figure \ref{numplot}, we would simply run  \\ \\
\verb"us={1.98066463867 - 0.63872734235 I,1.98066463867 + 0.63872734235 I," \\
\verb"-1.98066463867 + 0.63872734235 I,-1.98066463867 - 0.63872734235 I};"\\
\verb"vs={2.35231505474,-2.35231505474};"\\
\verb"r=Abs@Cxxo[16,4,16,2,4,2,us,vs]/Cooo[16,4,16,2,4,2]//FullForm" \\ \\
to obtain $r=0.976868082298$.

In table 2 and appendix D of \cite{paper1}, the scalar product between part of $\O_1$ and part of $\O_2$ needed to compute \verb"Cxxo" was written as a sum over all possible partitions of the roots involved (see equation (114) of \cite{paper1}). However, from a computational point of view it is much more convenient to write this scalar product using the new recursion relation for $SU(2)$ scalar products that we derived in equations (73), (77) and (78) of \cite{paper1}. This is in fact what we did to obtain the numerical results of this paper.  Despite this improvement, we were only able to consider cases where $\O_1$ had at most 14 roots. For example, it took
about 10 hours on a quad-core processor to compute the last point for $\alpha=1/3$ in figure \ref{numplot}.

\section{BPS 3-point function for $SU(3)$ sector} \la{BPSap}
For the setup in (\ref{Vacchoice2}) the structure constant between three BPS operators reads \cite{Lee:1998bxa}
\beqa
C_{123}^{\circ\circ\circ}= \frac{{J -j_1-j_2\choose J_1+J_2-j_1-j_2 }{J_1+J_2-j_1-j_2  \choose J_1-j_1 }{j_1+j_2  \choose j_1 }\sqrt{J j (J-j_1-j_2+j_3)} }{\sqrt{{J \choose J-J_3}{J-J_3 \choose J_2}{j \choose j_1+j_2}{j_1+j_2 \choose j_1} {J-j_1-j_2+j_3 \choose J_1+J_2-j_1-j_2}{J_1+J_2-j_1-j_2  \choose J_1-j_1}}}\;.
\la{su3bps}
\eeqa
In the classical limit considered in this paper this expression simplifies to (\ref{bpssimp}).

\section{More on back-reaction}\la{moreonbackreaction}

In this appendix we will consider the case where a root is added at infinity and will show analytically that back-reaction is very important. We will start in section \ref{O3S-} with a case where $C_{123}^{\bullet\bullet\circ}$ can be computed exactly without referring to coherent states. Then, in section \ref{weak_coupling_appendix} we will study in more generality the infinite root problem.

\subsection{A special BPS operator $\O_3=\Tr\(Z\bar X\)$}\la{O3S-}

In this section we will consider a specific example where $C_{123}^{\bullet\bullet\circ}$ can be computed exactly. We will then use it to draw some general conclusions about back-reaction. The case in mind is
\beq
\O_3=\Tr\(Z\bar X\)\nn
\eeq
and for simplicity, we take $\O_1$ and $\O_2$ to be in the $SU(2)$ sector. The point is that the Wick contraction of $\O_3$ with $\O_1$ flips one of the $X$ fields into a $Z$ field and therefore effectively acts as $S^+$. Now suppose we take $\O_1$ to be a Bethe eigenstate. Since Bethe eigenstates are highest weight states, they are annihilated by $S^+$, so $C_{123}^{\bullet\bullet\circ}=0$.

Now suppose that, instead, we take
\beq
\O_1={1\over\sqrt{J_3-J_1+2}}\[S^-,\bar\O_2\]\nn
\eeq
where $\O_2$ is represented by an exact Bethe eigenstate $|\psi_2\>$. Here $S^-$ flips one $Z$ field into an $X$ and is normalized such that it preserves the \textit{unit} norm of $|\psi_2\>$. We get that\footnote{$\sqrt{J_3-J_1+2}\,C_{123}^{\bullet\bullet\circ} =  \< \psi_2 | S^+ S^- |\psi_2\>=  \< \psi_2 | \[S^+ ,S^-\] |\psi_2\>= \< \psi_2 | S_z |\psi_2\>=(J_3-J_1+2)$}
\beq
C_{123}^{\bullet\bullet\circ}=\sqrt{J_3-J_1+2}\nn \, .
\eeq
For BPS operators, (\ref{su3bps}) reduces to
\beq
C_{123}^{\circ\circ\circ}=\sqrt{J_1(J_3+1)}\nn \, .
\eeq
In total we find that in the classical limit $C_{123}^{\bullet\bullet\circ}/C_{123}^{\circ\circ\circ}$ scales as $1/\sqrt{J}$, in agreement with the numerical experiments (\ref{infroots}). We see again that the result is very sensitive to what exactly one means by $\O_2\simeq\O_1$. In figure \ref{numplot2} we plotted the numerical results using the formulae from \cite{paper1} against the analytic prediction just derived. A perfect match is obtained.

\subsection{Root at infinity, more general scenario and any coupling}\la{weak_coupling_appendix}

We consider now a more general scenario which includes the setup of the previous subsection as a particular case and which is furthermore valid at any coupling. It illustrates very clearly the importance of back-reaction.

Consider three $SU(3)$ operators $\O_1,\O_2$ and $\O_3$ which are eigenstates of the exact dilatation operator.
We consider a small operator $\O_3$ and two large operator $\O_2$ and $\O_1 \simeq \bar \O_2$.

Suppose we now change the small BPS operator $\O_3$ by adding to it one more $\bar X$ field and one more $Z$ field. Schematically,
\beq
\widetilde\O_3=\O_3\cup\bar X\cup Z\nn \,.
\eeq
At the same time we keep the large operator $\O_2$ fixed. Due to R-charge conservation, the the $R$-charge of $\O_1$ must be changed accordingly. We can do that in many different ways by adding one more root to $\O_1$. If back-reaction would have been negligible, the result would not depend on where we add that extra root and the new three-point function would have been of the same order as the original one. As by now is clear from the  examples in the main text, that is not true. To see that in this general consideration, we choose to add the new root at infinity. That is, $\widetilde\O_1={\cal N}[S_X^-,\O_1]$, where $S_X^-$ is the $R$-charge rotation that rotates $Z$ field into  $X$ scalars and ${\cal N}=1/\sqrt{J_3-J_1}$ is a normalization factor of order $1/\sqrt J$, constructed such that the two point function of the new operator is still normalized to $1$.  Let us now consider the ratio
\beq\la{Calr}
R= C_{\,\widetilde 1 2\widetilde 3}/ C_{123}\nn \, .
\eeq
We will argue that $R$ is of order $1/\sqrt J$, which means that, as in the folded string example above, a small change in $\O_1$ and $\O_3$ leads to a different large $J$ scaling and back-reaction is very important.

The three point function is invariant under a global R-charge rotation of all operators. Hence we can trade the rotation of $\O_1$ by an action on $\O_3$ and $\O_2$. The action on $\O_2$ gives zero since it is an highest weight. The action on $\O_3$ yields a new small operator. On the other hand, the normalization factor $\mathcal{N} = O(1/\sqrt{J})$ remains and hence $R=O(1/\sqrt{J})$ as advocated.

A similar reasoning with more roots at infinity would lead to (\ref{infroots}).

\section{$\O_3= \Tr(Z^j)$  and (ignoring) mixing with double traces} \la{ex2}

In the main text we focused on operators $\O_3$ with $j_1+j_2 \neq 0$ scalars $\bar X,\bar Y$, and $j_3\neq 0$ scalars $Z$. The reason for considering this setup was pointed out in \cite{paper1}: in this setup all operators are Wick contracted to each other as depicted in figures \ref{3ptfunction} and \ref{HHL}.
When this is the case, a simple large $N$ counting shows us that we can ignore mixing of the single trace operators with double traces.

\begin{figure}[t]
\centering
\def\svgwidth{16cm}
\includegraphics{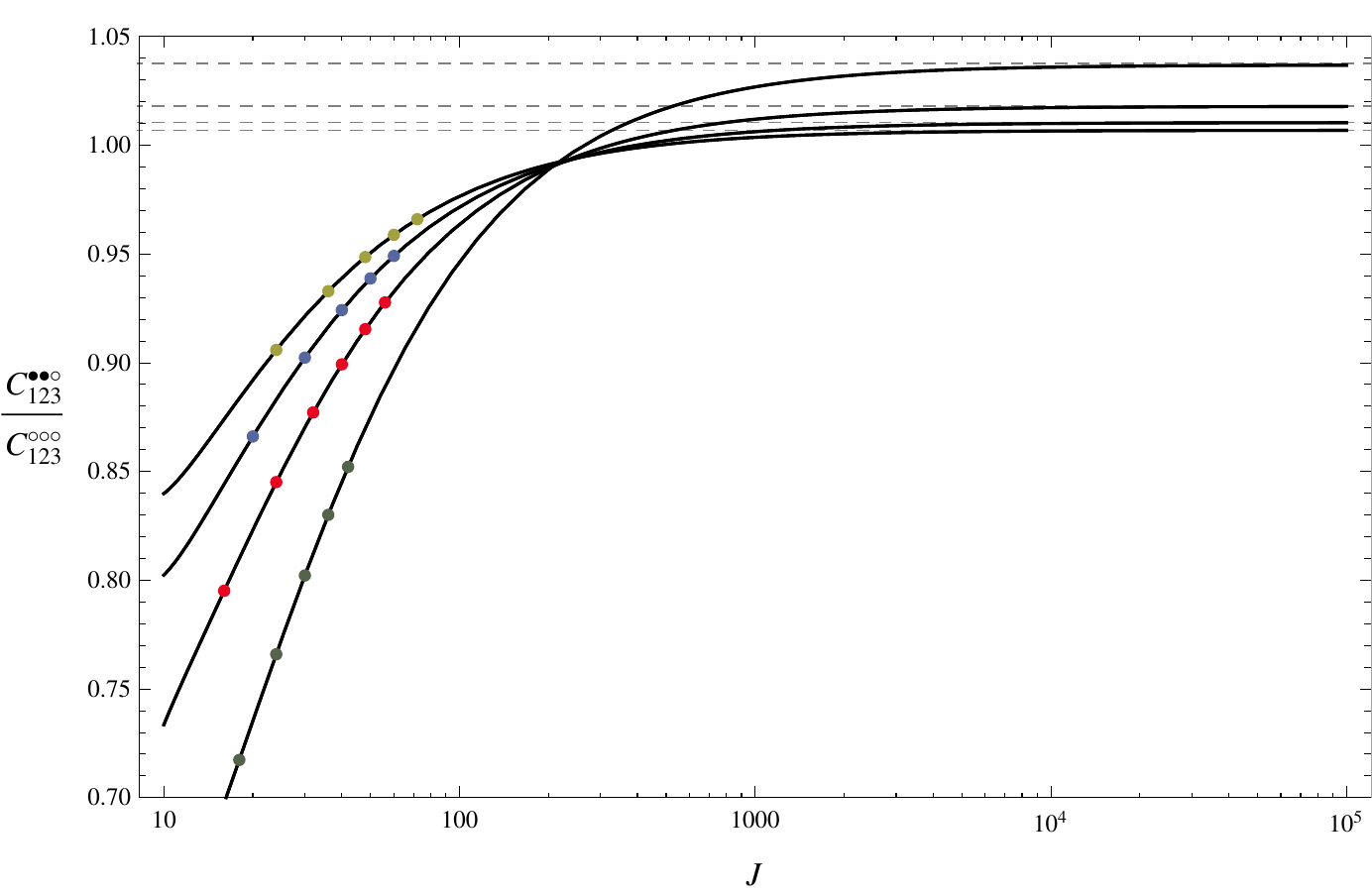}
\caption{Fits of the numerical data (black curves) compared to the analytical prediction (dashed gray lines)
for different values of the filling fraction of operator $\mathcal{O}_1$: $\alpha=1/3,1/4,1/5,1/6$ for $j=3$. No fine tuning is involved, see discussion at the end of section \ref{ex1}.
}\label{NumpD}
\end{figure}
Another setup corresponds to $j_1=j_2=0$,
\beq\O_3= \Tr(Z^j)\,.\eeq
In this case, the length of operator $\O_2$ is the sum of the lengths of the other two operators. In other words, there are now no  tree-level Wick contractions between $\O_1$ and $\O_3$. In this case to compute the structure constants we must also consider the mixing of operator $\O_2$ with double traces \cite{Beisert:2002bb}.\footnote{At the same time, we should mention that there is however considerable evidence in the literature that extremal correlators can be considered as an analytic continuation of non-extremal ones \cite{Rastelli}, see also \cite{new}.
} The setup in \cite{paper1} was chosen to avoid this complication.

On the other hand, the case $\O_3= \Tr(Z^j)$ is the case which has been most extensively studied in the literature recently at strong coupling
\cite{recentpapers1, recentpapers2, recentpapers3,Roiban:2010fe, common,Buchbinder:2010ek}. In all these computations, the mixing with multi-string states has not been taken into account. Let us ignore the mixing with double trace effect and see what we get in the classical limit at weak coupling. Basically, the hope is that the same effect is being forgotten at weak and strong coupling and hence ``cancels out" leaving a remainder quantity behind which can still be matched. Another plausible option is that this effect is negligible in the classical large $L$ limit

This seems indeed to be the case. From the point of view of the coherent state derivation, presented in section \ref{coherentderivation}, we can freely set $j_1=j_2=0$. Therefore, the weak coupling result (\ref{mainweak}) matches again the strong coupling result in the Frolov-Tseytlin limit, (\ref{mainstrong}).

Now, let us be more precise and face the kind of question which we already encountered in other examples described in the main text, see e.g. section \ref{ex1}. Namely, what exactly are we computing? Given a classical operator $\O_2$, what is the operator $\O_1 \simeq \bar\O_2$?
Operator $\O_1$ has $J_1$ scalars $X$ and total length $J$. Operator $\O_2$ has $J_1$ scalars $\bar X$  but a slightly larger length equal to $J+j$. Operator $\O_2$ is parametrized by some configuration of Bethe roots like the one in figure \ref{foldedcuts}. The position of the roots is uniquely fixed by a choice of mode numbers and by the length of the operator. Hence, a most natural choice for $\O_1$ is to take it to be the (conjugate of) operator whose Bethe roots have the same mode numbers but slightly smaller length.

To be more concrete, consider $\O_2$ to be the operator dual to the folded string introduced in section \ref{ex1}. Then $\O_1$ would be the same folded string with slightly smaller length. From the string point of view this would be a folded string with the same angular momentum in the plane identified by the scalar $X$ and with $j$ less units of angular momentum in the plane identified by the scalar $Z$.

We performed once more a numerical check of the coherent state prediction (\ref{mainweak})=(\ref{mainstrong}) against the results of \cite{paper1} when we put $j_1=0$ in the formulae of that paper. We find again a perfect match between the fits of the numerical data and the analytic prediction, as depicted in figure
\ref{NumpD}.

\end{document}

%% file: 3D3P.pdf_tex
%% Creator: Inkscape inkscape 0.48.0, www.inkscape.org
%% Accompanies image file '3D3P.pdf' (pdf, eps, ps)

\begingroup
  \makeatletter
  \providecommand\color[2][]{%
    \errmessage{(Inkscape) Color is used for the text in Inkscape, but the package 'color.sty' is not loaded}
    \renewcommand\color[2][]{}%
  }
  \providecommand\transparent[1]{%
    \errmessage{(Inkscape) Transparency is used (non-zero) for the text in Inkscape, but the package 'transparent.sty' is not loaded}
    \renewcommand\transparent[1]{}%
  }
  \providecommand\rotatebox[2]{#2}
  \ifx\svgwidth\undefined
    \setlength{\unitlength}{831.38212891pt}
  \else
    \setlength{\unitlength}{\svgwidth}
  \fi
  \global\let\svgwidth\undefined
  \makeatother
  \begin{picture}(1,1.02143621)%
    \put(0,0){\includegraphics[width=\unitlength]{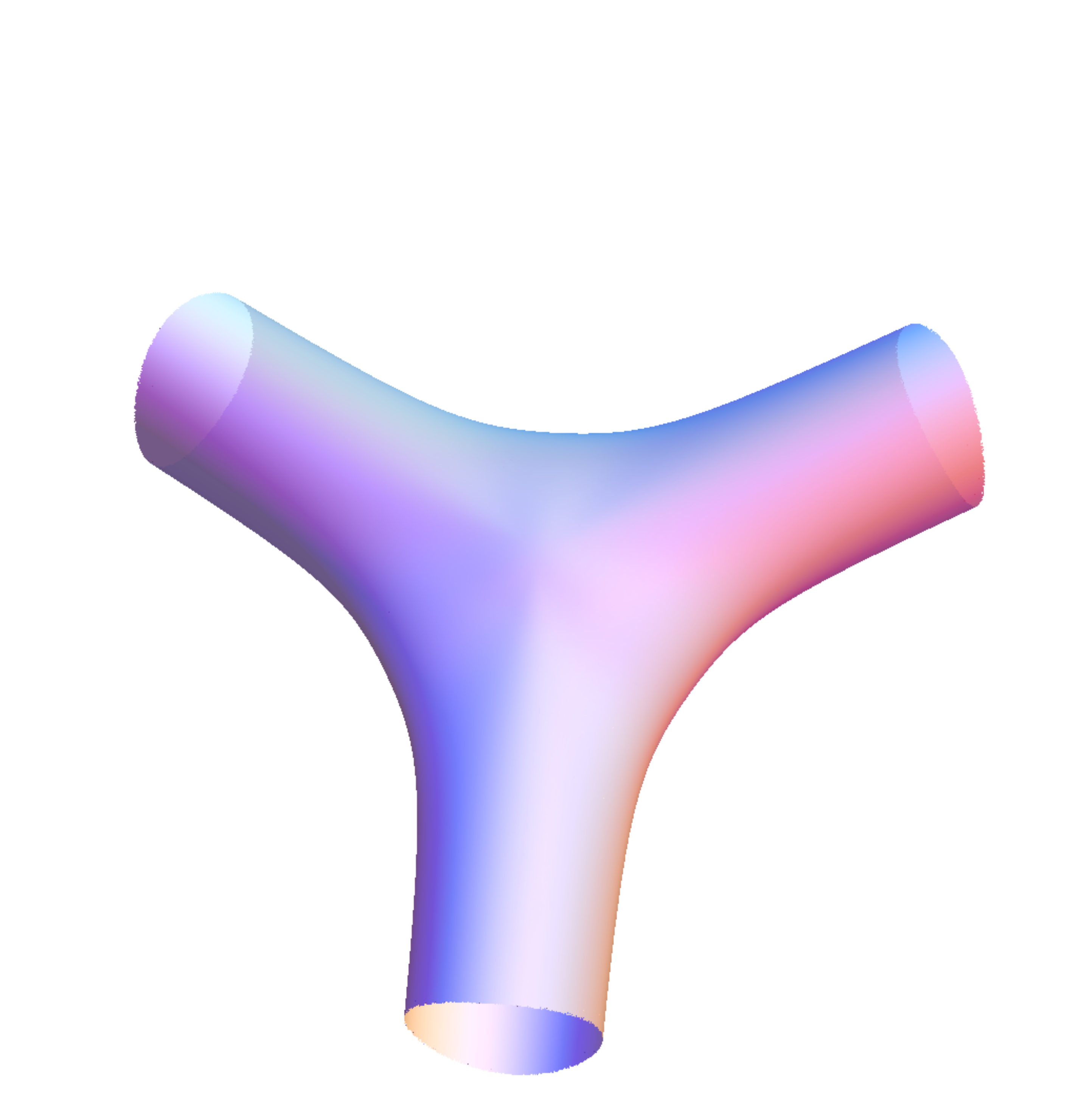}}%
    \put(0.1572487,0.02033345){\color[rgb]{0,0,0}\makebox(0,0)[lt]{\begin{minipage}{0.6113909\unitlength}\raggedright ${\cal O}_2=\sum\psi_2\ {\rm Tr}\left(...\bar Z...\color{red}\bar X\color{black}...\right)$\end{minipage}}}%
    \put(0.80753064,0.95677125){\color[rgb]{0,0,0}\rotatebox{-64.91174954}{\makebox(0,0)[lt]{\begin{minipage}{0.50114086\unitlength}\raggedright ${\cal O}_1=\sum\psi_1\ {\rm Tr}\left(... Z\dots\color{red}X\color{black}...\right)$\end{minipage}}}}%
    \put(-0.07282691,0.4516645){\color[rgb]{0,0,0}\rotatebox{60.14375821}{\makebox(0,0)[lt]{\begin{minipage}{0.5189655\unitlength}\raggedright ${\cal O}_3=\sum\psi_3\ {\rm Tr}\left(... Z\dots\color{red}\bar X\color{black}...\right)$\end{minipage}}}}%
    \put(0.77546276,0.07320671){\color[rgb]{0,0,0}\makebox(0,0)[lb]{\smash{$\la{Vacchoice}
\begin{array}{ccc}
& \text{vacuum} & \text{excitations} \\ \hline
\O_1 & Z & \color{red}X\color{black} \\
\O_2 & \bar Z &\color{red} \bar X\color{black} \\
\O_3 & Z &\color{red} \bar X\color{black}
\end{array}
$
}}}%
  \put(0,0){\includegraphics[width=\unitlength]{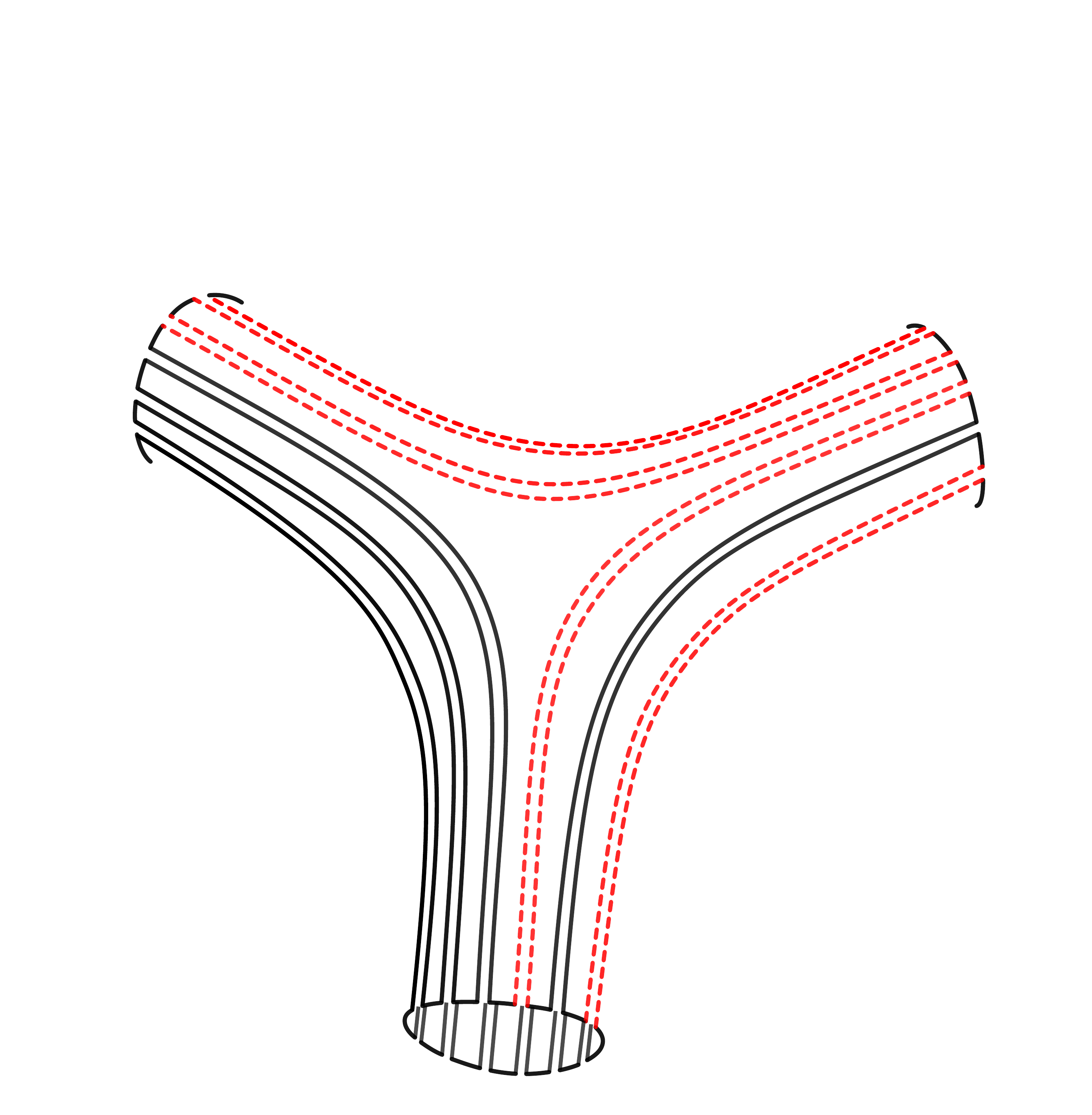}}%
  \end{picture}%
\endgroup

%% file: HHL.pdf_tex
%% Creator: Inkscape inkscape 0.48.0, www.inkscape.org
%% Accompanies image file 'HHL.pdf' (pdf, eps, ps)

\begingroup
  \makeatletter
  \providecommand\color[2][]{%
    \errmessage{(Inkscape) Color is used for the text in Inkscape, but the package 'color.sty' is not loaded}
    \renewcommand\color[2][]{}%
  }
  \providecommand\transparent[1]{%
    \errmessage{(Inkscape) Transparency is used (non-zero) for the text in Inkscape, but the package 'transparent.sty' is not loaded}
    \renewcommand\transparent[1]{}%
  }
  \providecommand\rotatebox[2]{#2}
  \ifx\svgwidth\undefined
    \setlength{\unitlength}{2287.14746094pt}
  \else
    \setlength{\unitlength}{\svgwidth}
  \fi
  \global\let\svgwidth\undefined
  \makeatother
  \begin{picture}(1,0.47429193)%
    \put(0,0){\includegraphics[width=\unitlength]{HHL}}%
    \put(-0.00039102,0.06971797){\color[rgb]{0,0,0}\makebox(0,0)[lb]{\smash{${\cal O}_2\!=\!\sum\psi_2{\rm Tr}\left(\bar Z...\color{red}\bar X\color{black}...\color{blue}\bar Y\color{black}...\right)$}}}%
    \put(0.31705651,0.26793032){\color[rgb]{0,0,0}\makebox(0,0)[lb]{\smash{${\cal O}_3=\sum{\rm Tr}\left( Z\color{red}\bar X\color{blue}\bar Y\color{black}\right)$}}}%
    \put(-0.00039102,0.46776841){\color[rgb]{0,0,0}\makebox(0,0)[lb]{\smash{${\cal O}_1\!=\!\sum\psi_1{\rm Tr}\left( Z...\color{red} X\color{black}...\color{blue} Y\color{black}...\right)$}}}%
    \put(0.11087819,0.01465204){\color[rgb]{0,0,0}\makebox(0,0)[lb]{\smash{$(a)$}}}%
    \put(0.71250097,0.01465204){\color[rgb]{0,0,0}\makebox(0,0)[lb]{\smash{$(b)$}}}%
    \put(0.53330925,0.08127676){\color[rgb]{0,0,0}\makebox(0,0)[lb]{\smash{${\cal O}_2\simeq{\bf u}_1{\bf u}_2{\bf u}_3\dots$}}}%
    \put(0.53330925,0.45484253){\color[rgb]{0,0,0}\makebox(0,0)[lb]{\smash{${\cal O}_1\simeq{\bf \bar u}_1{\bf \bar u}_2{\bf \bar u}_3\dots$}}}%
    \put(0.30780698,0.02238957){\color[rgb]{0,0,0}\makebox(0,0)[lb]{\smash{Large $J$ simplification}}}%
    \put(0.8975373,0.26933472){\color[rgb]{0,0,0}\makebox(0,0)[lb]{\smash{${\cal O}_3$}}}%
  \end{picture}%
\endgroup

%% file: Slice.pdf_tex
%% Creator: Inkscape inkscape 0.48.0, www.inkscape.org
%% Accompanies image file 'Slice.pdf' (pdf, eps, ps)

\begingroup
  \makeatletter
  \providecommand\color[2][]{%
    \errmessage{(Inkscape) Color is used for the text in Inkscape, but the package 'color.sty' is not loaded}
    \renewcommand\color[2][]{}%
  }
  \providecommand\transparent[1]{%
    \errmessage{(Inkscape) Transparency is used (non-zero) for the text in Inkscape, but the package 'transparent.sty' is not loaded}
    \renewcommand\transparent[1]{}%
  }
  \providecommand\rotatebox[2]{#2}
  \ifx\svgwidth\undefined
    \setlength{\unitlength}{576.8pt}
  \else
    \setlength{\unitlength}{\svgwidth}
  \fi
  \global\let\svgwidth\undefined
  \makeatother
  \begin{picture}(1,0.67428974)%
    \put(0,0){\includegraphics[width=\unitlength]{Slice.pdf}}%
    \put(0.82628216,0.11352518){\color[rgb]{0,0,0}\makebox(0,0)[lb]{\smash{$\O_1$}}}%
    \put(0.57940291,0.05616534){\color[rgb]{0,0,0}\makebox(0,0)[lb]{\smash{$\O_3$}}}%
    \put(0.23071729,0.23924433){\color[rgb]{0,0,0}\makebox(0,0)[lb]{\smash{$\O_2$}}}%
    \put(0.46698059,0.56733574){\color[rgb]{0,0,0}\rotatebox{42.82194881}{\makebox(0,0)[lb]{\smash{$\tau_e=0$}}}}%
  \end{picture}%
\endgroup

%% file: backreactions.pdf_tex
%% Creator: Inkscape inkscape 0.48.0, www.inkscape.org
%% Accompanies image file 'backreactions.pdf' (pdf, eps, ps)

\begingroup
  \makeatletter
  \providecommand\color[2][]{%
    \errmessage{(Inkscape) Color is used for the text in Inkscape, but the package 'color.sty' is not loaded}
    \renewcommand\color[2][]{}%
  }
  \providecommand\transparent[1]{%
    \errmessage{(Inkscape) Transparency is used (non-zero) for the text in Inkscape, but the package 'transparent.sty' is not loaded}
    \renewcommand\transparent[1]{}%
  }
  \providecommand\rotatebox[2]{#2}
  \ifx\svgwidth\undefined
    \setlength{\unitlength}{382.80883789pt}
  \else
    \setlength{\unitlength}{\svgwidth}
  \fi
  \global\let\svgwidth\undefined
  \makeatother
  \begin{picture}(1,0.26201845)%
    \put(0,0){\includegraphics[width=\unitlength]{backreactions}}%
    \put(0.09097132,0.00661888){\color[rgb]{0.50196078,0.50196078,0.50196078}\makebox(0,0)[lb]{\smash{$\text{(A)}$}}}%
    \put(0.46713819,0.00661888){\color[rgb]{0.50196078,0.50196078,0.50196078}\makebox(0,0)[lb]{\smash{$\text{(B)}$}}}%
    \put(0.82240689,0.00661888){\color[rgb]{0.50196078,0.50196078,0.50196078}\makebox(0,0)[lb]{\smash{$\text{(C)}$}}}%
  \end{picture}%
\endgroup

%% file: One_loop.pdf_tex
%% Creator: Inkscape inkscape 0.48.0, www.inkscape.org
%% Accompanies image file 'One_loop.pdf' (pdf, eps, ps)

\begingroup
  \makeatletter
  \providecommand\color[2][]{%
    \errmessage{(Inkscape) Color is used for the text in Inkscape, but the package 'color.sty' is not loaded}
    \renewcommand\color[2][]{}%
  }
  \providecommand\transparent[1]{%
    \errmessage{(Inkscape) Transparency is used (non-zero) for the text in Inkscape, but the package 'transparent.sty' is not loaded}
    \renewcommand\transparent[1]{}%
  }
  \providecommand\rotatebox[2]{#2}
  \ifx\svgwidth\undefined
    \setlength{\unitlength}{812.47094727pt}
  \else
    \setlength{\unitlength}{\svgwidth}
  \fi
  \global\let\svgwidth\undefined
  \makeatother
  \begin{picture}(1,1.11597656)%
    \put(0,0){\includegraphics[width=\unitlength]{One_loop.pdf}}%
    \put(0.53449313,0.44035369){\color[rgb]{0,0,0}\makebox(0,0)[lb]{\smash{$H$}}}%
    \put(0.273231,1.09877742){\color[rgb]{0,0,0}\makebox(0,0)[lb]{\smash{${\cal O}_1$}}}%
    \put(0.25353797,0.01369176){\color[rgb]{0,0,0}\makebox(0,0)[lb]{\smash{${\cal O}_2$}}}%
    \put(0.85220583,0.57100465){\color[rgb]{0,0,0}\makebox(0,0)[lb]{\smash{${\cal O}_3$}}}%
  \end{picture}%
\endgroup